\documentclass[journal=jacsat,manuscript=article]{achemso}
\usepackage{achemso}
\setkeys{acs}{maxauthors = 0}
\usepackage[version=3]{mhchem}
\usepackage[english]{babel}
\usepackage{filecontents}
\usepackage{makecell}
\usepackage{graphicx}
\usepackage{courier}
\usepackage{amssymb,amsmath,amsthm,mathrsfs,comment,afterpage,accents}
\usepackage{mathtools}
\usepackage{braket}
\usepackage{mathrsfs}
\usepackage{courier}
\usepackage{color}
\usepackage{subdepth}
\usepackage{tablefootnote}
\usepackage{hyperref}
\usepackage[capitalize]{cleveref}
\usepackage{multirow}
\usepackage[linesnumbered,ruled,vlined]{algorithm2e}
\setkeys{acs}{articletitle = true}
\makeatletter
\SectionNumbersOn
\def\@dotsep{4.5}
\usepackage{cancel}
\makeatother
\usepackage{dcolumn}
\usepackage{bm}
\newcommand\mat\mathbf

\newcommand{\insertrev}[1]{{\textcolor{black} {#1}}}
\newcommand{\insertrevtwo}[1]{{\textcolor{black} {#1}}}
\newcommand{\reprev}[2]{{\textcolor{red}{{}}}{\textcolor{black}{#2}}}

\title{
Utilizing Essential Symmetry Breaking in Auxiliary-Field Quantum Monte Carlo:
Application to the Spin Gaps of the \ce{C36} Fullerene and an Iron Porphyrin Model Complex
}

\author{Joonho Lee}
\email{linusjoonho@gmail.com}
\affiliation{
Department of Chemistry, Columbia University, New York, New York 10027, USA
}
\author{Fionn D. Malone}
\affiliation{Quantum Simulations Group, Lawrence Livermore National Laboratory, 7000 East Avenue, Livermore, CA, 94551 USA.}
\author{Miguel A. Morales}
\affiliation{Quantum Simulations Group, Lawrence Livermore National Laboratory, 7000 East Avenue, Livermore, CA, 94551 USA.}

\begin{document}
\maketitle
\begin{abstract}
We present three distinct examples where phaseless auxiliary-field Quantum Monte Carlo (ph-AFQMC)
can be reliably performed 
with a single-determinant trial wavefunction with {\it essential} symmetry breaking.
\insertrev{Essential symmetry breaking was
first introduced by Lee and Head-Gordon [{\it Phys. Chem. Chem. Phys.}, 2019, {\textbf {21}}, 4763-4778].}
We utilized essential \insertrev{complex and} time-reversal symmetry breaking with ph-AFQMC to compute the triplet-singlet energy gap in the TS12 set.
We found statistically better performance of ph-AFQMC with complex-restricted orbitals than with spin-unrestricted orbitals.
We then showed the utilization of essential spin symmetry breaking when computing the singlet-triplet gap of a known biradicaloid, \ce{C36}.
ph-AFQMC with spin-unrestricted Hartree-Fock (ph-AFQMC+UHF) fails catastrophically even with spin-projection and predicts no biradicaloid character. 
With approximate \insertrev{Brueckner} orbitals obtained from regularized orbital-optimized second-order M{\o}ller-Plesset perturbation theory ($\kappa$-OOMP2), 
ph-AFQMC quantitatively captures strong biradicaloid character of \ce{C36}. Lastly, we applied ph-AFQMC to the computation of the quintet-triplet gap in a model iron porphyrin complex 
where brute-force methods with a small active space fail to capture the triplet ground state. We show unambiguously that neither triplet nor quintet is strongly correlated using UHF, $\kappa$-OOMP2, and coupled-cluster with singles and doubles (CCSD) performed on UHF and $\kappa$-OOMP2 orbitals. There is no essential symmetry breaking in this problem.
By virtue of this, we were able to perform UHF+ph-AFQMC reliably with a cc-pVTZ basis set and predicted a triplet ground state for this model geometry.
The largest ph-AFQMC in this work correlated 186 electrons in 956 orbitals. Our work highlights the utility, scalability, and accuracy of ph-AFQMC with a
single determinant trial wavefunction with essential symmetry breaking for systems mainly dominated by dynamical correlation with little static correlation.
\end{abstract}
{\it Introduction} Spin energy gaps are important for characterizing electronic properties of molecules. They are often used as a parameter for determining the thermodynamic favorability of photocatalytic processes\cite{smith2010singlet}. Within the theoretical and computational quantum chemistry community, these gaps are useful in determining whether a given molecule exhibits strong correlation \cite{Rajca1994, Abe2013}. When the singlet-triplet gap of a given molecule is small (typically less than 10 kcal/mol), one may conclude that the molecule is biradicaloid \cite{Abe2013}. The accurate computation of singlet-triplet gaps of biradicaloids has been a challenging task in electronic structure theory \cite{Lee2019b}. This is in part due to the fact that singlet biradicaloids require a balanced treatment between strong and weak correlation. In terms of strong correlation, one has to treat at least two electrons in two orbitals (2e, 2o) beyond typical perturbation theory. At the same time, weak correlation out of this active space is important for quantitative accuracy. Although the separation between strong and weak correlation is often unclear and it is ambiguous to identify an active space, typical approaches are to apply a simple active space method to account for strong correlation and subsequently second-order perturbation theory to incorporate the remaining correlation effect. Popular approaches include second-order complete active-space perturbation theory (CASPT2) and second-order $N$-electron valence perturbation theory (NEVPT2) on top of a small active space CAS self-consistent field (CASSCF) reference state\cite{Andersson1990,Angeli2001}.

Recently, auxiliary-field quantum Monte Carlo (AFQMC) has received great attention in the {\it ab-initio} electronic structure community\citep{Motta2019}. 
It was initially developed for simulating the Hubbard model in the condensed matter physics community\cite{zhang_cpmc}, but has been further developed for general {\it ab-initio} systems\cite{Zhang_phaseless,al2006auxiliary,motta_back_prop,motta_forces}.
For molecular systems, it has been applied to bond dissociations\cite{Purwanto2008,Purwanto2015}, singlet-triplet gaps of biradicaloids\cite{Shee2019}, dipole bound anions\cite{hao2018accurate}, simple transition metal complexes\cite{Al-Saidi2006,shee2019achieving} and finite-temperature systems\cite{zhang_ftafqmc_99,liu2018ab}.
For solids, it has been applied to the uniform electron gas model\cite{lee_2019_UEG} as well as simple real solids such as boron nitride\cite{motta_kpoint}, hydrogen chains\cite{motta_hydrogen}, and nickel oxide\cite{zhang_nio}.
This broad range of applications was possible due to its relatively low cost compared to other popular many-body methods such as coupled-cluster with singles and doubles (CCSD). The computational bottleneck in AFQMC is either the propagation of walkers or the local energy evaluation depending on the choice of algorithms and discretization schemes\cite{suewattana2007phaseless,malone_isdf,motta_thc}. It should be noted that the existing algorithms and discretization schemes all fall into $\mathcal{O}(N^3)$ to $\mathcal{O}(N^4)$ scaling with system size $N$, which is favorable compared to the $\mathcal{O}(N^6)$ scaling of CCSD\cite{bartlett2007coupled}. 
This economical cost has allowed for one to perform AFQMC calculations without any active space restrictions even for medium-sized systems that are beyond the scope of conventional CCSD.

One of the more pressing challenges in AFQMC is the choice of trial wavefunctions. AFQMC is a projector QMC method and is similar in spirit to more commonly used diffusion MC (DMC)\cite{foulkes_dmc_review}. Namely, we obtain the ground state wavefunction using a projection method, 
\begin{equation}
|\Psi_0\rangle 
\propto
\lim_{\tau\rightarrow \infty}    
\exp{\left(-\tau \hat{H}\right)} |\Phi_0\rangle
= 
\lim_{\tau\rightarrow \infty}    
|\Psi(\tau)\rangle,
\label{eq:exact}
\end{equation}
where $\tau$ denotes the imaginary time, $\hat{H}$ is the Hamiltonian, 
$|\Psi_0\rangle$ is the exact ground state of $\hat{H}$,
and $|\Phi_0\rangle$ is an initial wavefunction satisfying $\langle\Phi_0|\Psi_0\rangle \ne 0$. 
\insertrev{The initial wavefunction $|\Phi_0\rangle$ can differ from the trial wavefunction $|\Psi_T\rangle$, we will assume $|\Phi_0\rangle = |\Phi_T\rangle$ unless mentioned otherwise.}
Based on this, it can in principle obtain the exact ground state of a given Hamiltonian, but it runs into the infamous fermionic sign problem. Similar to the fixed-node approximations in DMC, it is necessary to introduce an approximation into the algorithm to remove the sign problem. This approximation is called the phaseless approximation in AFQMC (i.e., ph-AFQMC).
In ph-AFQMC, one specifies a trial wavefunction that enforces a phase constraint so that the fermionic sign problem (or phase problem) no longer occurs. This is achieved by removing the phase in the overlap ratio factor used in the importance sampling:
\begin{equation}
S_n(\tau, \Delta\tau) = \frac{\langle
    \Psi_T | \psi_n(\tau+\Delta\tau)
    \rangle}{
    \langle
    \Psi_T | \psi_n(\tau)
    \rangle},
 \label{eq:ovl}
\end{equation}
where $\Delta \tau$ is the imaginary time step and $|\psi_n(\tau)\rangle$ is the wavefunction of the $n$-th walker at time $\tau$.
The phaseless approximation introduces uncontrollable biases into the resulting ph-AFQMC energy.
When practicing ph-AFQMC calculations, there needs to be great care in the choice of the trial wavefunction. 
In the recent paper by Shee and co-workers, the computation of singlet-triplet energy gaps over a variety of prototypical biradicaloids was carried out using ph-AFQMC\cite{Shee2019}. A simple spin-unrestricted Hartree-Fock single determinant state (UHF) in conjunction with a simple spin-projection\cite{Purwanto2008} was found to be effective for the systems considered in their work. \insertrev{Such a simple spin-projection is performed by setting the initial wavefunction $|\Phi_0\rangle$ to be a spin-pure determinant
using spin-restricted HF (RHF) or UHF natural orbitals.}
In some systems, UHF exhibited artificial spin-contamination and therefore they used spin-unrestricted Kohn-Sham density functional theory (UKS) for those systems.

In the present study, we investigate systems where UHF wavefunctions are qualitatively so wrong that the subsequent ph-AFQMC with spin-projection is quantitatively incorrect.
For problems within the scope of single-reference methods, we argue that \insertrev{Brueckner} orbitals define an optimal set of orbitals\cite{nesbet1958brueckner,Dykstra1977} and are a uniquely well-defined choice for the ph-AFQMC trial wavefunction. \reprev{They are an optimal choice in the sense that there is no trivial orbital rotation out of a determinant made of those orbitals.}{Any occupied-virtual rotation applied to the
starting wavefunction is redundant since the starting orbitals already encode this by construction. 
Subsequent imaginary-time propagation of walkers incorporates excitations higher than singles from the starting wavvefunction.}
Obtaining exact \insertrevtwo{Brueckner} orbitals is practically not feasible so we resort to using approximate ones \cite{Handy1989} either from orbital-optimized CC with doubles \cite{Krylov1998,Sherrill1998} or from regularized orbital-optimized M{\o}ller-Plesset perturbation theory ($\kappa$-OOMP2) \cite{Lee2018, Lee2019a, Lee2019b}. 
We note that the use of \insertrev{Brueckner} orbitals \insertrev{from CC with doubles} has been previously explored in the context of real-space DMC \cite{deible2016exploration}.

By optimizing orbitals in the presence of the regularized MP2 correlation energy, $\kappa$-OOMP2 removes {\it artificial} symmetry breaking and retains only {\it essential} symmetry breaking in a single determinant \cite{Lee2019a}. 
\insertrev{The regularized MP2 ($\kappa$-MP2) total energy is given as
\begin{equation}
E_{\kappa\text{-MP2}} = E_\text{HF}-\frac14 \sum_{ijab}\frac{|\langle ij||ab\rangle|^2}{\Delta_{ij}^{ab}} \left(1-e^{-\kappa\Delta_{ij}^{ab}}\right)^2
\label{eq:kmp2}
\end{equation}
It has a regularization damping function applied to the usual MP2 correlation energy and thereby it does not exhibit any
singularities or ill-defined behavior observed in canonical MP2\cite{Lee2018}.
When $\kappa \rightarrow 0$, it recovers the HF limit while for $\kappa \rightarrow \infty$ it recovers the unregularized MP2.
In $\kappa$-OOMP2, we seek a stationary point of \cref{eq:kmp2} with respect to orbital rotation parameters.
Interested readers are referred to refs. \citenum{Lee2019b}, \citenum{Lee2018}, and \citenum{Lee2019a} for implementation details.}
It has a favorable fifth-order scaling just like canonical MP2.
{\it Artificial} symmetry breaking is best exemplified by spin-symmetry breaking in closed-shell molecules. For instance, in \ce{C60}, a complex generalized HF (cGHF) solution was found\cite{Jimenez-Hoyos2014} and this solution was characterized to be artificial later on\cite{Lee2019a}. 
Essential symmetry breaking is {\it essential} because without it a single determinant wavefunction is qualitatively wrong. There are many examples for essential symmetry breaking ranging from bond dissociations to singlet biradicaloids. 
\insertrev{In summary, we resort to $\kappa$-OOMP2 for distinguishing {\it artificial} and {\it essential} symmetry breaking. Namely, when $\kappa$-OOMP2 orbitals break certain symmetries those are considered {\it essential}. On the other hand, any symmetry breaking that exists in HF orbitals but not in $\kappa$-OOMP2 orbitals is {\it artificial}.
One can determine whether a given symmetry breaking is {\it artificial} or {\it essential} by 
performing a $\kappa$-OOMP2 calculation starting from a broken-symmetry HF determinant.
Ideally, one should scan over $\kappa \in[0.0,4.0]$ as described in ref. \citenum{Lee2019a}.
For computational efficiency, in this work, we perform $\kappa$-OOMP2 at its optimal $\kappa$ value, 1.45\cite{Lee2018},
and characterize the HF symmetry breaking.}

\insertrev{In this work, we examine the efficacy of essential symmetry breaking in choosing a trial wavefunction for ph-AFQMC.} 
\insertrev{First, we will investigate prototypical systems where complex and time-reversal symmetry breaking is essential and show that it is statistically better to use a spin-restricted single determinant with complex and time-reversal symmetry breaking (i.e., complex, restricted (cR) orbitals) than that with spin-unrestriction.
These systems are where both HF and $\kappa$-OOMP2 exhibit genuine cR solutions and therefore the corresponding symmetry breaking is essential symmetry breaking.}
Furthermore, \ce{C36}, a known biradicaloid \cite{Fowler1999,Fowler1999a,Jagadeesh1999,Aihara1999,Lan-FengYuan2000,Slanina2000,Ito2000,Varganov2002,Stuck2011}, will be investigated with minimally spin-contaminated spin-unrestricted orbitals from $\kappa$-OOMP2. 
In \ce{C36}, spin-polarization is {\it essential} symmetry breaking. 
This is where 
\insertrev{spin-unrestricted $\kappa$-OOMP2 (i.e., $\kappa$-UOOMP2)} orbitals provide nearly optimal single-reference trial wavefunctions for ph-AFQMC. \insertrev{In contrast, UHF exhibits massive spin-polarization and subsequently UHF+ph-AFQMC is ill-behaved.}
Lastly, we will also study a model iron porphyrin complex which has been a topic of controversy between two selected configuration interaction methods\cite{smith2017cheap,levine2019casscf}.
UHF exhibits artificial symmetry breaking, which could be removed by $\kappa$-OOMP2. 
We will discuss how ph-AFQMC performs for this model transition metal system with single-reference trial wavefunctions.
\insertrev{The extent of artificial symmetry breaking was found to be small and therefore even UHF+ph-AFQMC performs qualitatively correctly.}
The important message of this paper is that ph-AFQMC combined with a broken-symmetry single determinant trial can provide accurate energetics when the underlying symmetry breaking is only essential.

\insertrev{We note that the cost for obtaining $\kappa$-OOMP2 orbitals scales as $\mathcal O(N^5)$, which is asymptotically more expensive than
subsequently performing ph-AFQMC with those orbitals. With tensor hypercontraction (THC) techniques\cite{Hohenstein2012}, this cost can be reduced to $\mathcal O(N^4)$\cite{lee_thc}, which is asymptotically
comparable to that of canonical ph-AFQMC. Since ph-AFQMC cost can be made to be $\mathcal O(N^3)$ with THC\cite{malone_isdf}, $\kappa$-OOMP2 is asymptotically more expensive even with THC.
However, we note that $\kappa$-OOMP2 has not been the actual computational bottleneck for systems considered \insertrevtwo{in this manuscript}.
}

\insertrevtwo{{\it Computational Details}
All deterministic calculations were performed with a development version of Q-Chem\cite{Shao2015}
and all QMC calculations were done with a development version of QMCPACK \cite{qmcpack,kent2020qmcpack}.
When running AFQMC, we used PySCF\cite{PYSCF} to generate two-electron integrals using Q-Chem's molecular orbital coefficients.
Implementational details of cRHF and $\kappa$-OOMP2 are given in ref. \citenum{Lee2019b} and ref. \citenum{Lee2018}, respectively.
The TS12 geometries are are all taken from ref. \citenum{Lee2019b}.
The \ce{C36} D$_\text{6h}$ geometry was taken from ref. \citenum{Lee2019a}, which is optimized at the level of BLYP/6-31G(d).
The Fe(P) geometry was taken from ref. \citenum{Groenhof2005} which is OPBE/TZP optimized geometry for $M_S=1$.
More computational details such as raw data relevant to the examples studied in this work and characterization of the UHF solutions found for Fe(P) are provided in the Supplementary Materials.}

\insertrev{{\it Essential Complex and Time-Reversal Symmetry Breaking}} We first present the study of the TS12 set which includes 12 atoms and diatomic molecules whose ground state is a triplet\cite{Lee2019b, lee2019kohn}.
\insertrev{An important feature of this data set is that complex and time-reversal symmetry breaking is essential.} Hence, the use of complex, restricted (cR) orbitals produces qualitatively better results when compared to spin-unrestricted (U) and spin-restricted (R) orbitals. 
This is due to the fact that a two-electron determinant made of a doubly occupied complex orbital $\xi = \eta + i\bar{\eta}$ (where $\eta$ and $\bar{\eta}$ are real orbitals) can describe two open-shell electrons\cite{Small2015a, Lee2019b, lee2019kohn}.
Thus, a cRHF trial wavefunction for these systems is a simple alternative to a multi-determinant trial wavefunction capable of describing the open-shell nature of the singlet ground state. 

\begin{table}[h!]
  \centering
  \begin{tabular}{c|r|r|r|r|r|r|r}
 &   Expt.  &  \shortstack{RHF\\+ph-AFQMC} &  \shortstack{UHF\\+ph-AFQMC} &  \shortstack{cRHF\\+ph-AFQMC}  &   RMP2  &   UMP2 &   cRMP2   \\\Xhline{3\arrayrulewidth}
\ce{C}  & 29.14 &   -2.5(2)  &   1.6(1)  &   -1.5(1)   & 13.85 & -13.58 & 1.36  \\\hline
\ce{NF}  & 34.32 &   1.7(4)  &   1.3(4)  &   -2.6(4)   & 10.99 & -17.23 & -1.7  \\\hline
\ce{NH}  & 35.93 &   2.0(2)  &   2.6(2)  &   -0.1(2)   & 15.9 & -17.29 & 0.59  \\\hline
\ce{NO-}  & 17.3 &   4.0(5)  &   6.6(5)  &   -0.1(5)   & 5.53 & -7.74 & -0.72  \\\hline
\ce{O2}  & 22.64 &   3.2(5)  &   3.6(5)  &   -1.5(5)   & 6.15 & 2.72 & -2.26  \\\hline
\ce{O}  & 45.37 &   -7.6(3)  &   0.5(1)  &   -1.5(1)   & 19.71 & -22.1 & 0.65  \\\hline
\ce{PF}  & 20.27 &   3.6(4)  &   2.7(4)  &   -0.6(3)   & 10.8 & -9.06 & 0.94  \\\hline
\ce{PH}  & 21.9 &   3.4(2)  &   2.7(2)  &   -0.4(2)   & 11.66 & -10.17 & 0.91  \\\hline
\ce{S2}  & 13.44 &   3.0(4)  &   4.5(5)  &   -1.6(4)   & 4.48 & -5.01 & -1.7  \\\hline
\ce{S}  & 26.41 &   -2.2(3)  &   1.5(2)  &   -2.3(2)   & 14.21 & -12.19 & 1.43  \\\hline
\ce{Si}  & 18.01 &   -0.6(2)  &   1.5(1)  &   -2.2(1)   & 10.12 & -7.76 & 1.45  \\\hline
\ce{SO}  & 18.16 &   2.3(5)  &   3.2(5)  &   -2.5(4)   & 3.94 & -9.84 & -3.49  \\\Xhline{3\arrayrulewidth}
RMSD  &   N/A  &   3.4(1)  &   3.1(1)  &   1.7(1)   & 11.6 & 12.42 & 1.64  \\\hline
MAD  &   N/A  &   3.0(1)  &   2.7(1)  &   1.40(9)   & 10.61 & 11.22 & 1.43  \\\hline
MSD  &   N/A  &   0.9(1)  &   2.7(1)  &   -1.40(9)    & 10.61 & -10.77 & -0.21  \\\Xhline{3\arrayrulewidth}
  \end{tabular}
  \caption{
The experimental triplet-singlet gap $\Delta E_\text{T-S} (= E_S - E_T)$ (kcal/mol) of systems in TS12 and the deviation (kcal/mol) in 
$\Delta E_\text{T-S}$ obtained from ph-AFQMC with spin-restricted (RHF), spin-unrestricted (UHF), and complex, spin-restricted (cRHF) trial wavefunctions. RMSD stands for root-mean-square-deviation, MAD is mean absolute deviation, and MSD stands for mean-signed-deviation. N/A means not applicable.
\insertrev{MP2 results are taken from ref. \citenum{Lee2019b}.}
}
  \label{tab:ts12}
\end{table}

In \cref{tab:ts12}, we present the deviation in triplet-singlet energy gaps of ph-AFQMC with RHF, UHF, and cRHF trial wavefunctions. 
The $M_S=1$ state is calculated with UHF trial wavefunctions as they are nearly spin-pure\cite{Lee2019b}.
These energies are from the complete basis set (CBS) limit extrapolated energies using aug-cc-pVTZ\cite{Dunning1989} and aug-cc-pVQZ\cite{Dunning1989} correlation energies via Helgaker's two-point extrapolation formula\cite{helgaker1997basis}.
The energy gaps are obtained by taking energy differences between UHF+ph-AFQMC ($M_S=1$) and
R/U/cRHF+ph-AFQMC ($M_S=0$). It should be noted that UHF+ph-AFQMC calculations for $M_S=0$ were performed with RHF initial states to obtain spin-pure states along the imaginary time propagation (i.e., spin-projected ph-AFQMC).
\cref{tab:ts12} shows a remarkably small trial wavefunction dependence in ph-AFQMC.
In typical electronic structure methods such as MP2, the reference orbital dependence is quite large. For TS12, the MP2 gap changes as much as 20 kcal/mol depending on what orbitals one uses \insertrev{as presented in \cref{tab:ts12}.}
Despite the small trial wavefunction dependence, it is clear that cRHF trial wavefunction leads to the statistically most accurate results for this set as shown in \cref{tab:ts12}.
Another notable aspect of this result is that UHF+ph-AFQMC overestimates the gaps for all systems whereas cRHF+ph-AFQMC underestimates the gap in every system.  The singlet state energy is a little too high in UHF+ph-AFQMC while it is a little too low in cRHF+ph-AFQMC. RHF+ph-AFQMC is the best performing ph-AFQMC method in terms of MSD but individual data points are quantitatively far worse than those of UHF or cRHF+ph-AFQMC. For instance, the \ce{O} atom has an error of -7.6(3) kcal/mol from RHF+ph-AFQMC whereas UHF+ph-AFQMC predicts 0.5(1) kcal/mol and cRHF+ph-AFQMC predicts -1.5(1) kcal/mol errors.
In passing we note that the use of $\kappa$-OOMP2 orbitals in the trial wavefunction makes virtually no differences compared to the corresponding HF orbitals in this case.
\insertrev{Those numbers are reported in the Supplementary Materials.}

{\it Essential Spin Symmetry Breaking} The essential symmetry breaking in \ce{C36} is spin polarization and there is no \insertrev{genuine cRHF solution} \cite{Lee2019a}. In UHF with cc-pVDZ\cite{Dunning1989}, $\langle \hat{S}^2 \rangle$ values are 7.4 and 8.7 for $M_S=0$ and $M_S=1$ states, respectively. This spin contamination can be removed in ph-AFQMC by starting from spin-pure single determinant wavefunctions (i.e., spin-projection\cite{Purwanto2008}). Here, we form a spin-restricted determinant out of UHF natural orbitals and use it as an initial walker determinant. On the other hand, orbitals from $\kappa$-UOOMP2 exhibit 
$\langle \hat{S}^2 \rangle$ values of 1.1 and 2.1, respectively, for $M_S=0$ and $M_S=1$ states.
The triplet state is nearly spin-pure $\langle \hat{S}^2 \rangle \simeq 2.0$, which asserts the fact that the $M_S=1$ state is
well described by a single determinant.
On the contrary, the singlet state exhibits $\langle \hat{S}^2 \rangle$ of 1.1 which shows strong biradicaloid character.
This demonstrates that spin symmetry breaking is {\it essential} in this system and we will see how this affects the
accuracy of ph-AFQMC calculations. We note that all electrons are correlated in ph-AFQMC and $\kappa$-UOOMP2 calculations.

\begin{table}[h!]
  \centering
  \begin{tabular}{c|c|r|r|r}\hline
Method & Spin-Projection & \multicolumn{1}{c|}{$\Delta E_\text{S-T}$} & $\langle \hat{S}^2\rangle_{M_S = 0}$ & $\langle \hat{S}^2\rangle_{M_S = 1}$\\\hline
RHF & N/A & -20.83 & 0.0 & 2.0 \\ \hline
UHF & No & 26.64 & 7.4 & 8.7 \\ \hline
MRMP2\cite{Varganov2002} & N/A & 8.17 & 0.0 & 2.0\\ \hline
AP+$\kappa$-UOOMP2 & Yes & 9.22 & 0.0 & 2.1 \\ \hline
RHF+ph-AFQMC & N/A  & 3.5(8) & 0.0 & 2.0 \\ \hline
UHF+ph-AFQMC & Yes & 42.7(6) & 0.0 & 2.0 \\ \hline
$\kappa$-UOOMP2+ph-AFQMC & No &  7(1) & N/A & N/A \\ \hline
$\kappa$-UOOMP2+ph-AFQMC & Yes &  6.4(9) & 0.0 & 2.0 \\ \hline
  \end{tabular}
  \caption{
  The vertical singlet-triplet gap $\Delta E_\text{S-T}  (= E_T - E_S)$  (kcal/mol) and the expectation values of $\langle \hat{S}^2\rangle$ for  $M_S=0$ and $M_S=1$ of \ce{C36} from various methods.
 \insertrev{AP indicates Yamaguchi's approximate spin-projection.}
All but MRMP2 results were obtained with the cc-pVDZ basis set \cite{Dunning1989}.
MRMP2 results in ref. \citenum{Varganov2002} were obtained with a D$_\text{6h}$ geometry within the 6-31G(d) basis set \cite{Hehre1972}. 
N/A means not available.
 \insertrev{$\langle \hat{S}^2\rangle$ for $\kappa$-UOOMP2+ph-AFQMC without spin-projection is not available since the corresponding mixed estimator
 has not yet been implemented.}
}
  \label{tab:c36}
\end{table}

In \cref{tab:c36}, we present the vertical singlet-triplet gap of \ce{C36} computed via various methods. 
At the level of HF, RHF predicts the sign of the gap wrong and UHF predicts a large spin gap with massive spin contamination. 
The predicted HF gaps (-20.83 kcal/mol and 26.65 kcal/mol) are far away from that of \insertrevtwo{multi-reference MP2 (MRMP2)\cite{Varganov2002}} (8.17 kcal/mol). 
The MRMP2 calculation was performed on a CASSCF solution with a (2e,4o) active space which 
covers a very small fraction of the entire $\pi$-space. 
It is unclear whether this is a reliable reference value so 
ph-AFQMC will provide another highly accurate reference singlet-triplet gap for this molecule.
$\kappa$-OOMP2 in conjunction with Yamaguchi's spin projection\cite{Yamaguchi1988} predicts the gap of 9.22 kcal/mol. 
This is about 1 kcal/mol larger than the MRMP2 value.
With the RHF trial wavefunctions for both $M_S=0$ and $M_S=1$ states, ph-AFQMC predicts a gap of 3.5(8) kcal/mol. 
The triplet energy was found to be too low compared to other ph-AFQMC calculations. This led to a very small singlet-triplet gap.
The internal stability analysis indicates that the solution used here is stable, but it is possible that the $M_S=1$ ROHF solution is a local minimum and the global minimum solution may yield a better AFQMC result. ROHF tends to experience more local minima problems than UHF, and thus this can be the case here as well.

UHF+ph-AFQMC is far worse than RHF+ph-AFQMC and even UHF. The gap of 42.7(6) kcal/mol is too large to be considered to be biradicaloid and it is about 34 kcal/mol away from the MRMP2 and $\kappa$-UOOMP2 results.
With $\kappa$-UOOMP2 orbitals, ph-AFQMC predicts the gap of 7(1) kcal/mol and 6.4(9) kcal/mol, respectively, with or without spin-projection. These two energies are well within the \insertrev{statistical} error bar of each other.
The $\kappa$-UOOMP2+ph-AFQMC gap is almost within the \insertrev{statistical} error bar from MRMP2 and is small enough to show biradicaloid character of \ce{C36}. The spin-projected $\kappa$-UOOMP2 with cc-pVTZ gap is 8.46 kcal/mol. Therefore, we expect the basis set incompleteness error to be on the order of 1 kcal/mol. It will be interesting to revisit this problem with a larger basis set to examine the basis set incompleteness error.
These results highlight the utility of $\kappa$-OOMP2 in generating approximate \insertrev{Brueckner} orbitals which can be used to form a trial determinant for ph-AFQMC calculations.

{\it Artificial Spin Symmetry Breaking}
\begin{figure}[h!]
\includegraphics[scale=0.40]{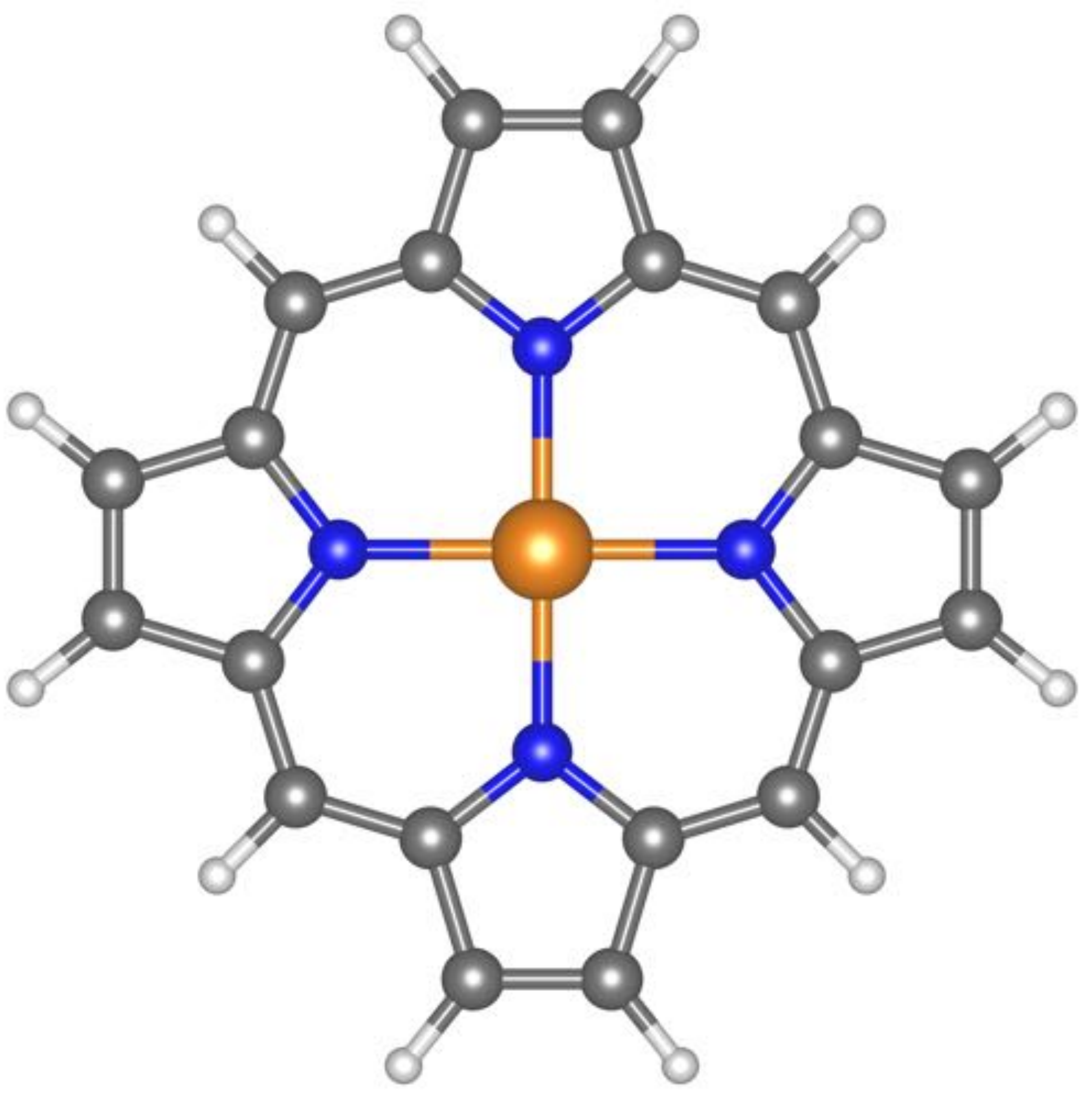}
\caption{
Molecular structure of the iron porphyrin (Fe(P)) model complex.
Cartesian coordinates were taken from ref. \citenum{Groenhof2005}. The color codes are as follows: Fe (orange), N (blue), C (gray), and H (white).
}
\label{fig:fep}
\end{figure}
Iron porphyrin complexes (shown in Figure \ref{fig:fep}) are abundant catalysts in numerous reactions that undergo in biological systems.
In particular, it plays a crucial role in the heme group in hemoglobin and myoglobin\cite{hori1980iron}.
Earlier density functional theory (DFT) studies indicate that its triplet and quintet states are of single-reference where as its singlet state may exhibit biradicaloid character\cite{Matsuzawa1995, Rovira1997,Groenhof2005, Radon2008, Radon2014}.
In addition to DFT, many wavefunction methods have been applied to the computation of the triplet and quintet energy gap. These largely include multi-reference (MR) methods like multi-reference MP2\cite{Choe1999}, CASPT2\cite{Pierloot2003,Phung2018}, second-order restricted active space peturbation theory (RASPT2)\cite{Vancoillie2011}, density matrix renormalization group configuration interaction (DMRG-CI)\cite{Olivares-Amaya2015}, DMRG-CASPT2\cite{Phung2016}, full configuration interaction QMC SCF (FCIQMC-SCF)\cite{LiManni2018,LiManni2019}, heat-bath configuration interaction SCF (HCISCF)\cite{smith2017cheap}, DMRG with pair DFT (DMRG-PDFT)\cite{Zhou2019}, and adaptively sampled CI SCF (ASCISCF)\cite{levine2019casscf}.
All of these methods require a specification of active space which can be as small as (8e, 11o) and as large as (44e, 44o). 
As shown in \cref{tab:fe}, This then gives a spectrum of the quintet-triplet gap ($\Delta E_\text{Q-T}  (= E_T - E_Q)$) from -13 kcal/mol (DMRG-CI) to 19.27(7) kcal/mol (ASCISCF).
Perhaps, the most surprising result in this broad spectrum is that two selected CI methods (SHCISCF and ACISCF) show a discrepancy on the order of 20 kcal/mol.
This is troublesome because the experimental spin energy gap is unavailable and the only available information is that the ground state is a triplet \cite{collman1975synthesis,dolphin1976synthesis,goff1977nuclear,lang1978mossbauer,kitagawa1979resonance,boyd1979paramagnetic,mispelter1980proton,strauss1985comparison}.
This broad spectrum does not necessarily indicate that the accuracy of MR methods differs within a given active space or one is better than others. 
In our view, the fact that one has to specify a small active space for computational feasibility is preventing us from the direct computation of this spin energy gap without neglecting dynamic correlation out of the active space.
In particular, there is a need for a method which does not resort to the cancellation of missing dynamic correlation out of the active space.

\begin{table}[h!]
  \centering
  \begin{tabular}{c|r|c|c|c}\Xhline{3\arrayrulewidth}
Method & $\Delta E_\text{Q-T}$ & Active space & Basis set & Relativistic \\ \Xhline{3\arrayrulewidth}
CASPT2\cite{Vancoillie2011} & 2.1 & 8e, 11o & ANO-RCC & DKH \\ \hline
CASPT2\cite{Vancoillie2011} & -0.9 & 16e, 15o & ANO-RCC & DKH \\ \hline
FCIQMC-SCF\cite{LiManni2018} & -3.1 & 32e, 34o & ANO-RCC & DKH \\ \hline
RASPT2\cite{Vancoillie2011} & 4.6 & 34e, 35o & ANO-RCC & DKH \\ \hline
DMRG-PDFT:ftPBE\cite{Zhou2019} & -0.7 & 34e, 35o & ANO-RCC & DKH \\ \hline
FCIQMC-SCF\cite{LiManni2019} & -4.4 & 40e, 38o & ANO-RCC & No \\ \hline
DMRG-CI\cite{Olivares-Amaya2015} & -13 & 44e, 44o & cc-pVDZ & No \\ \hline
SHCISCF\cite{smith2017cheap} & -1.9(7) & 44e, 44o & cc-pVDZ & No \\ \hline
ACISCF\cite{levine2019casscf} & 19.27(7) & 44e, 44o & cc-pVDZ & No \\ \Xhline{3\arrayrulewidth}
RCCSD(T)\cite{Pierloot2017} & 0.6 & frozen core (no 3s, 3p) & ANO-RCC & DKH \\ \hline
RCCSDTQ\cite{LiManni2019} & -4.8 & 40e, 38o & ANO-RCC & No \\ \hline
UHF & 26.6 & no active space & cc-pVDZ & No \\ \hline
$\kappa$-UOOMP2 & -1.5 & no active space & cc-pVDZ & No \\ \hline
UCCSD & 3.1 & no active space & cc-pVDZ & No \\ \hline
UCCSD(T) & -1.4 & no active space & cc-pVDZ & No \\ \hline
UCCSD:$\kappa$-UOOMP2 & 1.7 & no active space & cc-pVDZ & No \\ \hline
UHF+AFQMC & -1.7(5) & no active space & cc-pVDZ & No \\ \hline
ROHF+AFQMC & -3.4(6) & no active space & cc-pVDZ & No \\ \hline
$\kappa$-UOOMP2+AFQMC & -6.1(7) & no active space & cc-pVDZ & No \\ \hline
UHF & 27.2 & no active space & cc-pVTZ & No \\ \hline
$\kappa$-UOOMP2 & -3.4 & no active space & cc-pVTZ & No \\ \hline
UHF+AFQMC & \insertrev{-6.6(7)} 
& no active space & cc-pVTZ & No \\ \Xhline{3\arrayrulewidth}
\end{tabular}
  \caption{
  The vertical quintet-triplet gap ($\Delta E_\text{Q-T}  (= E_T - E_Q)$)  (kcal/mol) of the model iron porphyrin complex.
  DKH stands for the scalar relativistic correction via the Douglas-Kroll-Hess Hamiltonian\cite{douglas1974quantum}.
  The ANO-RCC basis set employs a contraction scheme that yields triple-zeta basis set quality \cite{widmark1990density,roos2004main,roos2005new}.
  UCCSD:$\kappa$-UOOMP2 indicates UCCSD on top of $\kappa$-UOOMP2 orbitals\cite{bertels2019third}.
}
  \label{tab:fe}
\end{table}

During the course of investigation, we discovered that neither triplet nor quintet is strongly correlated with the triplet geometry optimized with DFT as used in refs. \citenum{Groenhof2005,smith2017cheap,levine2019casscf}. 
This conclusion was drawn from several indications found in multiple single-reference (SR) methods such as UHF, spin-unrestricted coupled-cluster singles and doubles (UCCSD), UCCSD with perturbative triples (UCCSD(T)), $\kappa$-UOOMP2, and UCCSD on top of $\kappa$-UOOMP2 orbitals (UCCSD:$\kappa$-UOOMP2)\cite{bertels2019third}. UHF solutions exhibit $\langle \hat{S}^2\rangle$ of 4.01 and 7.82 for triplet and quintet states, respectively.
This apparent spin-contamination can be almost completely removed by $\kappa$-UOOMP2 which yields $\langle \hat{S}^2\rangle$ of 2.02 and 6.03, respectively. While the quintet-triplet energy gap is small, the spin contamination in the triplet state by the quintet state was found to be negligible. 
There is no {\it essential} symmetry breaking for this system in either spin state and UHF symmetry breaking is only {\it artificial}. Therefore, this problem should be in the reach of SR methods.
Furthermore, UCCSD calculations on top of spin-contaminated UHF solutions exhibit no substantial doubles amplitudes. The largest $T_2$ amplitudes are 0.0503 for triplet and 0.0511 for quintet. While UCCSD exhibits quite significant $T_1$ amplitudes (0.2455 for triplet and 0.2437 for quintet), even these amplitudes become small in UCCSD:$\kappa$-UOOMP2 (0.0944 for triplet and 0.0408 for quintet). This asserts the validity of $\kappa$-UOOMP2 as an approximate \insertrev{Brueckner} orbital method for this problem as well as the SR character of the problem.
While there has been no diagnosis of MR or SR character for this problem based on spin-symmetry breaking, previous CC studies with spin-restricted orbitals
indicated that this problem is mainly a dynamic correlation problem and does not necessarily require brute-force active space methods\cite{Pierloot2017,Phung2018,LiManni2019}.
Motivated by all these indications and also a recent discrepancy between SHCISCF and ACISCF, we employed UCCSD(T) and ph-AFQMC with all electrons correlated
within the cc-pVDZ basis set. This is far beyond the reach of active space methods since this corresponds to an active space of (186e, 439o). 

UHF yields a gap of 26.6 kcal/mol with a quintet ground state.
In order to obtain a triplet ground state, a reasonable correlation model is supposed to decrease this gap to a negative value.
In this sense, 19.27(7) kcal/mol of ACISCF is surprising because the correlation out of the active space is so significant that it seems to achieve not much of the
cancellation of dynamic correlation. By contrast, all other previous MR studies achieved either a triplet ground state or at least small enough gaps (less than 5 kcal/mol) by benefiting from the cancellation of dynamic correlation.
UCCSD and UCCSD(T) yield gaps of 3.1 kcal/mol and -1.4 kcal/mol. The correlation beyond doubles is responsible for obtaining a triplet ground state in this geometry.
We expect that the correlation beyond (T) may play some role in stabilizing the triplet state further by 1 kcal/mol or so as suggested in ref. \citenum{LiManni2019}.
$\kappa$-UOOMP2 yields a gap of -1.5 kcal/mol while UCCSD:$\kappa$-UOOMP2 shows a gap of 1.2 kcal/mol.
It is very likely that $\kappa$-UOOMP2 correlation model itself is insufficient to provide quantitative accuracy.
The difference between UCCSD and UCCSD:$\kappa$-UOOMP2 gaps is 1.4 kcal/mol, which is due to artificial symmetry breaking in UHF and is not negligible.

We then applied ph-AFQMC with UHF, ROHF, and $\kappa$-UOOMP2 trial wavefunctions.
In ph-AFQMC+UHF, spin-projection makes no improvement so ph-AFQMC itself seems to restore the underlying broken symmetry. 
The gap of ph-AFQMC+UHF is -1.7(5) kcal/mol which is within the error bar of UCCSD(T). The use of ROHF and $\kappa$-UOOMP2 trial wavefunctions leads to the gaps of -3.4(6) kcal/mol and -6.0(7) kcal/mol, respectively.
While the range of gap varies depending on what trial wavefunction one uses (just like CCSD gaps depending on the reference wavefunction), they all consistently predict a triplet ground state in this geometry. Lastly, we also performed the UHF+ph-AFQMC calculation with the cc-pVTZ basis set\cite{Dunning1989} while correlating all electrons.
This corresponds to an active space of (186e, 956o) and is beyond the scope of canonical CCSD and CCSD(T) assuming limited resources. 
This is also the largest AFQMC calculation done in this paper.
The UHF gap changed from 26.6 kcal/mol (cc-pVDZ) to 27.2 kcal/mol (cc-pVTZ), which suggests that the occupied orbitals are nearly converged to the basis set limit.
The $\kappa$-UOOMP2 gap changed by about 2 kcal/mol and predicts an increased gap of -3.4 kcal/mol.
The UHF+ph-AFQMC/cc-pVTZ gap is \insertrev{-6.6(7)} kcal/mol which shows a much larger gap than that of cc-pVDZ. 
This large negative ph-AFQMC gap provides enough margin for predicting a triplet ground state using the adiabatic quintet-triplet energy gap in the future.
All calculations presented in \cref{tab:fe} are based on a triplet DFT geometry and thus it will be important to revisit this problem with geometries that are optimized for 
each spin state with some reasonably accurate correlation models.

{\it Conclusions} In summary, we showed the utility of single-reference trial wavefunctions based on {\it essential} symmetry breaking when performing ph-AFQMC calculations.
We observed statistically better performance of ph-AFQMC when combined with complex, restricted orbitals than with spin-unrestricted orbitals in the TS12 set where \insertrev{complex and} time-reversal symmetry breaking is essential. We also showed a catastrophic failure of ph-AFQMC+UHF which could not be fixed by a simple spin-projection in the case of computing the singlet-triplet energy gap of \ce{C36}. An approximate \insertrev{Brueckner} orbital method, $\kappa$-UOOMP2, was shown to provide a set of qualitatively correct orbitals. ph-AFQMC+$\kappa$-UOOMP2 yielded a gap of 6.4(9) kcal/mol confirming the well-known biradicaloid character of \ce{C36}. Lastly, we showed strong evidence on the lack of multi-reference character in an iron porphyrin model complex. The UHF spin-symmetry breaking is only artificial and even with simple trial wavefunctions based on UHF, ROHF, and $\kappa$-UOOMP2, we observed consistently a triplet ground state. The examples and approach shown in this work highlight the usefulness of ph-AFQMC even with simple single-determinant trial wavefunctions. We believe that such a ph-AFQMC approach is most useful for systems which are mainly dominated by dynamic correlation and too large for canonical coupled-cluster methods to run.

{\it Acknowldegement}
We thank Sandeep Sharma for helpful discussion on the iron porphyrin complex.
J. L. thanks Martin Head-Gordon for his multi-year critical involvement in the development of $\kappa$-OOMP2 and Soojin Lee for encouragement.
This work was performed under the auspices of the U.S. Department of Energy
(DOE) by LLNL under Contract No. DE-AC52-07NA27344.  The work of F. D. M and M. A. M. was supported by 
the U.S. DOE, Office of Science, Basic Energy Sciences, Materials Sciences and
Engineering Division, as part of the Computational Materials Sciences Program
and Center for Predictive Simulation of Functional Materials (CPSFM).  Computing support for this work came from the LLNL Institutional Computing Grand Challenge program.

\section*{Table of Contents}
\begin{figure}[h!]
\includegraphics[scale=0.40]{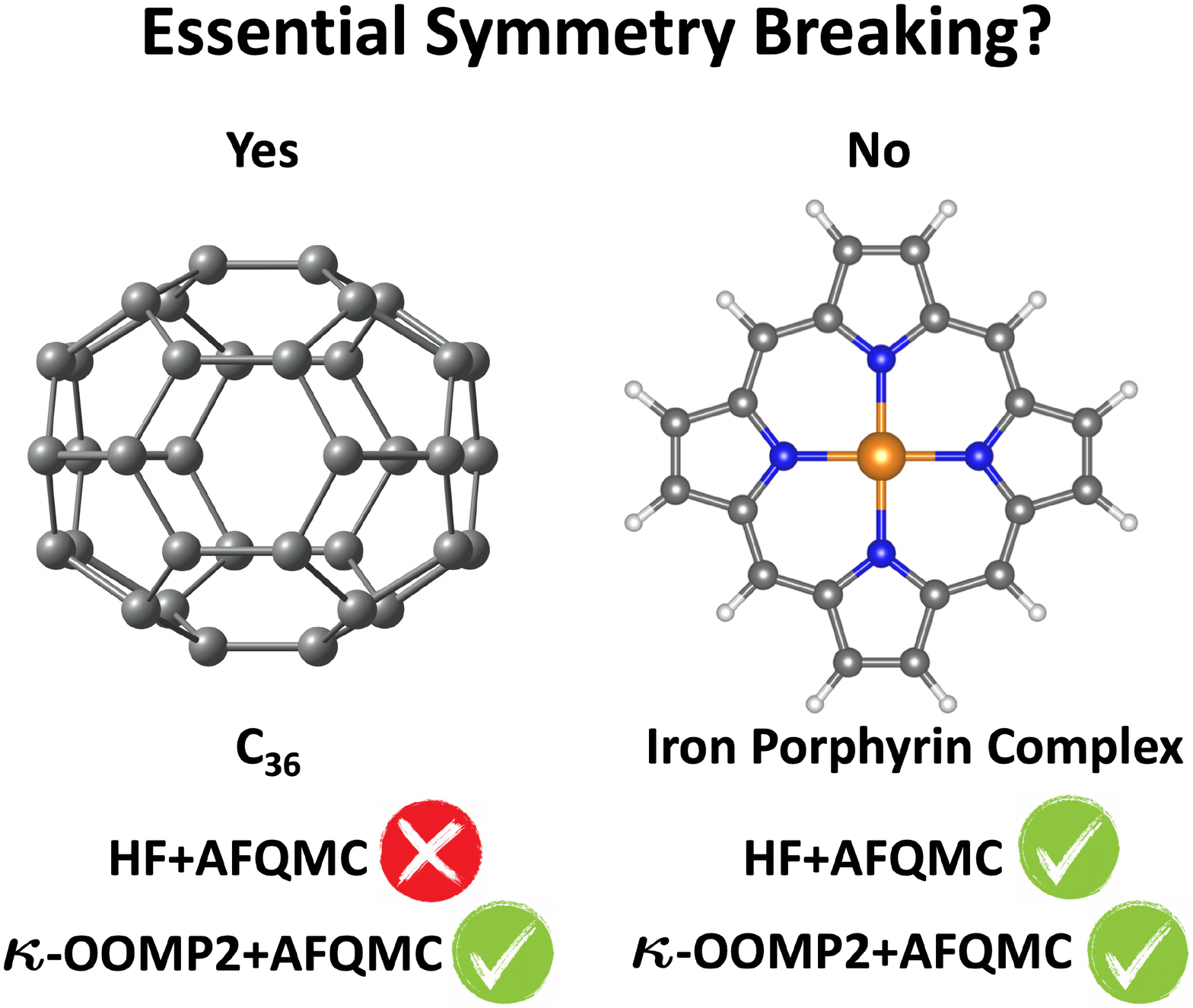}
\end{figure}
\bibliography{refs}

\providecommand{\latin}[1]{#1}
\makeatletter
\providecommand{\doi}
  {\begingroup\let\do\@makeother\dospecials
  \catcode`\{=1 \catcode`\}=2 \doi@aux}
\providecommand{\doi@aux}[1]{\endgroup\texttt{#1}}
\makeatother
\providecommand*\mcitethebibliography{\thebibliography}
\csname @ifundefined\endcsname{endmcitethebibliography}
  {\let\endmcitethebibliography\endthebibliography}{}
\begin{mcitethebibliography}{91}
\providecommand*\natexlab[1]{#1}
\providecommand*\mciteSetBstSublistMode[1]{}
\providecommand*\mciteSetBstMaxWidthForm[2]{}
\providecommand*\mciteBstWouldAddEndPuncttrue
  {\def\EndOfBibitem{\unskip.}}
\providecommand*\mciteBstWouldAddEndPunctfalse
  {\let\EndOfBibitem\relax}
\providecommand*\mciteSetBstMidEndSepPunct[3]{}
\providecommand*\mciteSetBstSublistLabelBeginEnd[3]{}
\providecommand*\EndOfBibitem{}
\mciteSetBstSublistMode{f}
\mciteSetBstMaxWidthForm{subitem}{(\alph{mcitesubitemcount})}
\mciteSetBstSublistLabelBeginEnd
  {\mcitemaxwidthsubitemform\space}
  {\relax}
  {\relax}

\bibitem[Smith and Michl(2010)Smith, and Michl]{smith2010singlet}
Smith,~M.~B.; Michl,~J. Singlet fission. \emph{Chem. Rev.} \textbf{2010},
  \emph{110}, 6891--6936\relax
\mciteBstWouldAddEndPuncttrue
\mciteSetBstMidEndSepPunct{\mcitedefaultmidpunct}
{\mcitedefaultendpunct}{\mcitedefaultseppunct}\relax
\EndOfBibitem
\bibitem[Rajca(1994)]{Rajca1994}
Rajca,~A. Organic Diradicals and Polyradicals: From Spin Coupling to Magnetism?
  \emph{Chem. Rev.} \textbf{1994}, \emph{94}, 871--893\relax
\mciteBstWouldAddEndPuncttrue
\mciteSetBstMidEndSepPunct{\mcitedefaultmidpunct}
{\mcitedefaultendpunct}{\mcitedefaultseppunct}\relax
\EndOfBibitem
\bibitem[Abe(2013)]{Abe2013}
Abe,~M. {Diradicals}. \emph{Chem. Rev.} \textbf{2013}, \emph{113},
  7011--7088\relax
\mciteBstWouldAddEndPuncttrue
\mciteSetBstMidEndSepPunct{\mcitedefaultmidpunct}
{\mcitedefaultendpunct}{\mcitedefaultseppunct}\relax
\EndOfBibitem
\bibitem[Lee and Head-Gordon(2019)Lee, and Head-Gordon]{Lee2019b}
Lee,~J.; Head-Gordon,~M. Two single-reference approaches to singlet
  biradicaloid problems: Complex, restricted orbitals and approximate
  spin-projection combined with regularized orbital-optimized M{\o}ller-Plesset
  perturbation theory. \emph{J. Chem. Phys.} \textbf{2019}, \emph{150},
  244106\relax
\mciteBstWouldAddEndPuncttrue
\mciteSetBstMidEndSepPunct{\mcitedefaultmidpunct}
{\mcitedefaultendpunct}{\mcitedefaultseppunct}\relax
\EndOfBibitem
\bibitem[Andersson \latin{et~al.}(1990)Andersson, Malmqvist, Roos, Sadlej, and
  Wolinski]{Andersson1990}
Andersson,~K.; Malmqvist,~P.~A.; Roos,~B.~O.; Sadlej,~A.~J.; Wolinski,~K.
  {Second-order perturbation theory with a CASSCF reference function}. \emph{J.
  Phys. Chem.} \textbf{1990}, \emph{94}, 5483--5488\relax
\mciteBstWouldAddEndPuncttrue
\mciteSetBstMidEndSepPunct{\mcitedefaultmidpunct}
{\mcitedefaultendpunct}{\mcitedefaultseppunct}\relax
\EndOfBibitem
\bibitem[Angeli \latin{et~al.}(2001)Angeli, Cimiraglia, Evangelisti, Leininger,
  and Malrieu]{Angeli2001}
Angeli,~C.; Cimiraglia,~R.; Evangelisti,~S.; Leininger,~T.; Malrieu,~J.-P.
  Introduction of n-electron valence states for multireference perturbation
  theory. \emph{J. Chem. Phys.} \textbf{2001}, \emph{114}, 10252--10264\relax
\mciteBstWouldAddEndPuncttrue
\mciteSetBstMidEndSepPunct{\mcitedefaultmidpunct}
{\mcitedefaultendpunct}{\mcitedefaultseppunct}\relax
\EndOfBibitem
\bibitem[Motta and Zhang(2018)Motta, and Zhang]{Motta2019}
Motta,~M.; Zhang,~S. Ab initio computations of molecular systems by the
  auxiliary-field quantum Monte Carlo method. \emph{WIREs Comput. Mol. Sci.}
  \textbf{2018}, \emph{8}, e1364\relax
\mciteBstWouldAddEndPuncttrue
\mciteSetBstMidEndSepPunct{\mcitedefaultmidpunct}
{\mcitedefaultendpunct}{\mcitedefaultseppunct}\relax
\EndOfBibitem
\bibitem[Zhang \latin{et~al.}(1997)Zhang, Carlson, and Gubernatis]{zhang_cpmc}
Zhang,~S.; Carlson,~J.; Gubernatis,~J.~E. Constrained path Monte Carlo method
  for fermion ground states. \emph{Phys. Rev. B} \textbf{1997}, \emph{55},
  7464\relax
\mciteBstWouldAddEndPuncttrue
\mciteSetBstMidEndSepPunct{\mcitedefaultmidpunct}
{\mcitedefaultendpunct}{\mcitedefaultseppunct}\relax
\EndOfBibitem
\bibitem[Zhang and Krakauer(2003)Zhang, and Krakauer]{Zhang_phaseless}
Zhang,~S.; Krakauer,~H. Quantum Monte Carlo Method using Phase-Free Random
  Walks with Slater Determinants. \emph{Phys. Rev. Lett.} \textbf{2003},
  \emph{90}, 136401\relax
\mciteBstWouldAddEndPuncttrue
\mciteSetBstMidEndSepPunct{\mcitedefaultmidpunct}
{\mcitedefaultendpunct}{\mcitedefaultseppunct}\relax
\EndOfBibitem
\bibitem[Al-Saidi \latin{et~al.}(2006)Al-Saidi, Zhang, and
  Krakauer]{al2006auxiliary}
Al-Saidi,~W.; Zhang,~S.; Krakauer,~H. Auxiliary-field quantum Monte Carlo
  calculations of molecular systems with a Gaussian basis. \emph{J. Chem.
  Phys.} \textbf{2006}, \emph{124}, 224101\relax
\mciteBstWouldAddEndPuncttrue
\mciteSetBstMidEndSepPunct{\mcitedefaultmidpunct}
{\mcitedefaultendpunct}{\mcitedefaultseppunct}\relax
\EndOfBibitem
\bibitem[Motta and Zhang(2017)Motta, and Zhang]{motta_back_prop}
Motta,~M.; Zhang,~S. Computation of Ground-State Properties in Molecular
  Systems: Back-Propagation with Auxiliary-Field Quantum Monte Carlo. \emph{J.
  Chem. Theory Comput.} \textbf{2017}, \emph{13}, 5367\relax
\mciteBstWouldAddEndPuncttrue
\mciteSetBstMidEndSepPunct{\mcitedefaultmidpunct}
{\mcitedefaultendpunct}{\mcitedefaultseppunct}\relax
\EndOfBibitem
\bibitem[Motta and Zhang(2018)Motta, and Zhang]{motta_forces}
Motta,~M.; Zhang,~S. Communication: Calculation of interatomic forces and
  optimization of molecular geometry with auxiliary-field quantum Monte Carlo.
  \emph{J. Chem. Phys.} \textbf{2018}, \emph{148}, 181101\relax
\mciteBstWouldAddEndPuncttrue
\mciteSetBstMidEndSepPunct{\mcitedefaultmidpunct}
{\mcitedefaultendpunct}{\mcitedefaultseppunct}\relax
\EndOfBibitem
\bibitem[Purwanto \latin{et~al.}(2008)Purwanto, Al-Saidi, Krakauer, and
  Zhang]{Purwanto2008}
Purwanto,~W.; Al-Saidi,~W.~A.; Krakauer,~H.; Zhang,~S. Eliminating spin
  contamination in auxiliary-field quantum Monte Carlo: Realistic potential
  energy curve of \ce{F2}. \emph{J. Chem. Phys.} \textbf{2008}, \emph{128},
  114309\relax
\mciteBstWouldAddEndPuncttrue
\mciteSetBstMidEndSepPunct{\mcitedefaultmidpunct}
{\mcitedefaultendpunct}{\mcitedefaultseppunct}\relax
\EndOfBibitem
\bibitem[Purwanto \latin{et~al.}(2015)Purwanto, Zhang, and
  Krakauer]{Purwanto2015}
Purwanto,~W.; Zhang,~S.; Krakauer,~H. {An auxiliary-field quantum Monte Carlo
  study of the chromium dimer}. \emph{J. Chem. Phys.} \textbf{2015},
  \emph{142}, 064302\relax
\mciteBstWouldAddEndPuncttrue
\mciteSetBstMidEndSepPunct{\mcitedefaultmidpunct}
{\mcitedefaultendpunct}{\mcitedefaultseppunct}\relax
\EndOfBibitem
\bibitem[Shee \latin{et~al.}(2019)Shee, Arthur, Zhang, Reichman, and
  Friesner]{Shee2019}
Shee,~J.; Arthur,~E.~J.; Zhang,~S.; Reichman,~D.~R.; Friesner,~R.~A.
  Singlet–Triplet Energy Gaps of Organic Biradicals and Polyacenes with
  Auxiliary-Field Quantum Monte Carlo. \emph{J. Chem. Theory Comput.}
  \textbf{2019}, \emph{15}, 4924--4932\relax
\mciteBstWouldAddEndPuncttrue
\mciteSetBstMidEndSepPunct{\mcitedefaultmidpunct}
{\mcitedefaultendpunct}{\mcitedefaultseppunct}\relax
\EndOfBibitem
\bibitem[Hao \latin{et~al.}(2018)Hao, Shee, Upadhyay, Ataca, Jordan, and
  Rubenstein]{hao2018accurate}
Hao,~H.; Shee,~J.; Upadhyay,~S.; Ataca,~C.; Jordan,~K.~D.; Rubenstein,~B.~M.
  Accurate Predictions of Electron Binding Energies of Dipole-Bound Anions via
  Quantum Monte Carlo Methods. \emph{J. Phys. Chem. Lett.} \textbf{2018},
  \emph{9}, 6185--6190\relax
\mciteBstWouldAddEndPuncttrue
\mciteSetBstMidEndSepPunct{\mcitedefaultmidpunct}
{\mcitedefaultendpunct}{\mcitedefaultseppunct}\relax
\EndOfBibitem
\bibitem[Al-Saidi \latin{et~al.}(2006)Al-Saidi, Krakauer, and
  Zhang]{Al-Saidi2006}
Al-Saidi,~W.~A.; Krakauer,~H.; Zhang,~S. {Auxiliary-field quantum Monte Carlo
  study of TiO and MnO molecules}. \emph{Phys. Rev. B} \textbf{2006},
  \emph{73}, 075103\relax
\mciteBstWouldAddEndPuncttrue
\mciteSetBstMidEndSepPunct{\mcitedefaultmidpunct}
{\mcitedefaultendpunct}{\mcitedefaultseppunct}\relax
\EndOfBibitem
\bibitem[Shee \latin{et~al.}(2019)Shee, Rudshteyn, Arthur, Zhang, Reichman, and
  Friesner]{shee2019achieving}
Shee,~J.; Rudshteyn,~B.; Arthur,~E.~J.; Zhang,~S.; Reichman,~D.~R.;
  Friesner,~R.~A. On Achieving High Accuracy in Quantum Chemical Calculations
  of 3 d Transition Metal-Containing Systems: A Comparison of Auxiliary-Field
  Quantum Monte Carlo with Coupled Cluster, Density Functional Theory, and
  Experiment for Diatomic Molecules. \emph{J. Chem. Theory Comput.}
  \textbf{2019}, \emph{15}, 2346--2358\relax
\mciteBstWouldAddEndPuncttrue
\mciteSetBstMidEndSepPunct{\mcitedefaultmidpunct}
{\mcitedefaultendpunct}{\mcitedefaultseppunct}\relax
\EndOfBibitem
\bibitem[Zhang(1999)]{zhang_ftafqmc_99}
Zhang,~S. Finite-Temperature Monte Carlo Calculations for Systems with
  Fermions. \emph{Phys. Rev. Lett.} \textbf{1999}, \emph{83}, 2777--2780\relax
\mciteBstWouldAddEndPuncttrue
\mciteSetBstMidEndSepPunct{\mcitedefaultmidpunct}
{\mcitedefaultendpunct}{\mcitedefaultseppunct}\relax
\EndOfBibitem
\bibitem[Liu \latin{et~al.}(2018)Liu, Cho, and Rubenstein]{liu2018ab}
Liu,~Y.; Cho,~M.; Rubenstein,~B. Ab initio finite temperature auxiliary field
  quantum monte carlo. \emph{J. Chem. Theory Comput.} \textbf{2018}, \emph{14},
  4722--4732\relax
\mciteBstWouldAddEndPuncttrue
\mciteSetBstMidEndSepPunct{\mcitedefaultmidpunct}
{\mcitedefaultendpunct}{\mcitedefaultseppunct}\relax
\EndOfBibitem
\bibitem[Lee \latin{et~al.}(2019)Lee, Malone, and Morales]{lee_2019_UEG}
Lee,~J.; Malone,~F.~D.; Morales,~M.~A. An auxiliary-Field quantum Monte Carlo
  perspective on the ground state of the dense uniform electron gas: An
  investigation with Hartree-Fock trial wavefunctions. \emph{J. Chem. Phys.}
  \textbf{2019}, \emph{151}, 064122\relax
\mciteBstWouldAddEndPuncttrue
\mciteSetBstMidEndSepPunct{\mcitedefaultmidpunct}
{\mcitedefaultendpunct}{\mcitedefaultseppunct}\relax
\EndOfBibitem
\bibitem[Motta \latin{et~al.}(2019)Motta, Zhang, and Chan]{motta_kpoint}
Motta,~M.; Zhang,~S.; Chan,~G. K.-L. Hamiltonian symmetries in auxiliary-field
  quantum Monte Carlo calculations for electronic structure. \emph{Phys. Rev.
  B} \textbf{2019}, \emph{100}, 045127\relax
\mciteBstWouldAddEndPuncttrue
\mciteSetBstMidEndSepPunct{\mcitedefaultmidpunct}
{\mcitedefaultendpunct}{\mcitedefaultseppunct}\relax
\EndOfBibitem
\bibitem[Motta \latin{et~al.}(2017)Motta, Ceperley, Chan, Gomez, Gull, Guo,
  Jim{\'e}nez-Hoyos, Lan, Li, Ma, \latin{et~al.} others]{motta_hydrogen}
Motta,~M.; Ceperley,~D.~M.; Chan,~G. K.-L.; Gomez,~J.~A.; Gull,~E.; Guo,~S.;
  Jim{\'e}nez-Hoyos,~C.~A.; Lan,~T.~N.; Li,~J.; Ma,~F., \latin{et~al.}  Towards
  the solution of the many-electron problem in real materials: equation of
  state of the hydrogen chain with state-of-the-art many-body methods.
  \emph{Phys. Rev. X} \textbf{2017}, \emph{7}, 031059\relax
\mciteBstWouldAddEndPuncttrue
\mciteSetBstMidEndSepPunct{\mcitedefaultmidpunct}
{\mcitedefaultendpunct}{\mcitedefaultseppunct}\relax
\EndOfBibitem
\bibitem[Zhang \latin{et~al.}(2018)Zhang, Malone, and Morales]{zhang_nio}
Zhang,~S.; Malone,~F.~D.; Morales,~M.~A. Auxiliary-field quantum Monte Carlo
  calculations of the structural properties of nickel oxide. \emph{J. Chem.
  Phys.} \textbf{2018}, \emph{149}, 164102\relax
\mciteBstWouldAddEndPuncttrue
\mciteSetBstMidEndSepPunct{\mcitedefaultmidpunct}
{\mcitedefaultendpunct}{\mcitedefaultseppunct}\relax
\EndOfBibitem
\bibitem[Suewattana \latin{et~al.}(2007)Suewattana, Purwanto, Zhang, Krakauer,
  and Walter]{suewattana2007phaseless}
Suewattana,~M.; Purwanto,~W.; Zhang,~S.; Krakauer,~H.; Walter,~E.~J. Phaseless
  auxiliary-field quantum Monte Carlo calculations with plane waves and
  pseudopotentials: Applications to atoms and molecules. \emph{Phys. Rev. B}
  \textbf{2007}, \emph{75}, 245123\relax
\mciteBstWouldAddEndPuncttrue
\mciteSetBstMidEndSepPunct{\mcitedefaultmidpunct}
{\mcitedefaultendpunct}{\mcitedefaultseppunct}\relax
\EndOfBibitem
\bibitem[Malone \latin{et~al.}(2019)Malone, Zhang, and Morales]{malone_isdf}
Malone,~F.~D.; Zhang,~S.; Morales,~M.~A. Overcoming the Memory Bottleneck in
  Auxiliary Field Quantum Monte Carlo Simulations with Interpolative Separable
  Density Fitting. \emph{J. Chem. Theory. Comput.} \textbf{2019}, \emph{15},
  256\relax
\mciteBstWouldAddEndPuncttrue
\mciteSetBstMidEndSepPunct{\mcitedefaultmidpunct}
{\mcitedefaultendpunct}{\mcitedefaultseppunct}\relax
\EndOfBibitem
\bibitem[Motta \latin{et~al.}(2019)Motta, Shee, Zhang, and Chan]{motta_thc}
Motta,~M.; Shee,~J.; Zhang,~S.; Chan,~G. K.-L. Efficient Ab Initio
  Auxiliary-Field Quantum Monte Carlo Calculations in Gaussian Bases via
  Low-Rank Tensor Decomposition. \emph{J. Chem. Theory Comput.} \textbf{2019},
  \emph{15}, 3510--3521\relax
\mciteBstWouldAddEndPuncttrue
\mciteSetBstMidEndSepPunct{\mcitedefaultmidpunct}
{\mcitedefaultendpunct}{\mcitedefaultseppunct}\relax
\EndOfBibitem
\bibitem[Bartlett and Musia{\l}(2007)Bartlett, and
  Musia{\l}]{bartlett2007coupled}
Bartlett,~R.~J.; Musia{\l},~M. Coupled-cluster theory in quantum chemistry.
  \emph{Rev. Mod. Phys.} \textbf{2007}, \emph{79}, 291\relax
\mciteBstWouldAddEndPuncttrue
\mciteSetBstMidEndSepPunct{\mcitedefaultmidpunct}
{\mcitedefaultendpunct}{\mcitedefaultseppunct}\relax
\EndOfBibitem
\bibitem[Foulkes \latin{et~al.}(2001)Foulkes, Mitas, Needs, and
  Rajagopal]{foulkes_dmc_review}
Foulkes,~W. M.~C.; Mitas,~L.; Needs,~R.~J.; Rajagopal,~G. Quantum Monte Carlo
  simulations of solids. \emph{Rev. Mod. Phys.} \textbf{2001}, \emph{73},
  33--83\relax
\mciteBstWouldAddEndPuncttrue
\mciteSetBstMidEndSepPunct{\mcitedefaultmidpunct}
{\mcitedefaultendpunct}{\mcitedefaultseppunct}\relax
\EndOfBibitem
\bibitem[Nesbet(1958)]{nesbet1958brueckner}
Nesbet,~R. Brueckner's Theory and the Method of Superposition of
  Configurations. \emph{Phys. Rev.} \textbf{1958}, \emph{109}, 1632\relax
\mciteBstWouldAddEndPuncttrue
\mciteSetBstMidEndSepPunct{\mcitedefaultmidpunct}
{\mcitedefaultendpunct}{\mcitedefaultseppunct}\relax
\EndOfBibitem
\bibitem[Dykstra(1977)]{Dykstra1977}
Dykstra,~C.~E. {An examination of the Brueckner condition for the selection of
  molecular orbitals in correlated wavefunctions}. \emph{Chem. Phys. Lett.}
  \textbf{1977}, \emph{45}, 466--469\relax
\mciteBstWouldAddEndPuncttrue
\mciteSetBstMidEndSepPunct{\mcitedefaultmidpunct}
{\mcitedefaultendpunct}{\mcitedefaultseppunct}\relax
\EndOfBibitem
\bibitem[Handy \latin{et~al.}(1989)Handy, Pople, Head-Gordon, Raghavachari, and
  Trucks]{Handy1989}
Handy,~N.~C.; Pople,~J.~A.; Head-Gordon,~M.; Raghavachari,~K.; Trucks,~G.~W.
  {Size-consistent Brueckner theory limited to double substitutions}.
  \emph{Chem. Phys. Lett.} \textbf{1989}, \emph{164}, 185--192\relax
\mciteBstWouldAddEndPuncttrue
\mciteSetBstMidEndSepPunct{\mcitedefaultmidpunct}
{\mcitedefaultendpunct}{\mcitedefaultseppunct}\relax
\EndOfBibitem
\bibitem[Krylov \latin{et~al.}(1998)Krylov, Sherrill, Byrd, and
  Head-Gordon]{Krylov1998}
Krylov,~A.~I.; Sherrill,~C.~D.; Byrd,~E. F.~C.; Head-Gordon,~M.
  {Size-consistent wave functions for nondynamical correlation energy: The
  valence active space optimized orbital coupled-cluster doubles model}.
  \emph{J. Chem. Phys.} \textbf{1998}, \emph{109}, 10669\relax
\mciteBstWouldAddEndPuncttrue
\mciteSetBstMidEndSepPunct{\mcitedefaultmidpunct}
{\mcitedefaultendpunct}{\mcitedefaultseppunct}\relax
\EndOfBibitem
\bibitem[Sherrill \latin{et~al.}(1998)Sherrill, Krylov, Byrd, and
  Head-Gordon]{Sherrill1998}
Sherrill,~C.~D.; Krylov,~A.~I.; Byrd,~E. F.~C.; Head-Gordon,~M. {Energies and
  analytic gradients for a coupled-cluster doubles model using variational
  Brueckner orbitals: Application to symmetry breaking in \ce{O4+}}. \emph{J.
  Chem. Phys.} \textbf{1998}, \emph{109}, 4171\relax
\mciteBstWouldAddEndPuncttrue
\mciteSetBstMidEndSepPunct{\mcitedefaultmidpunct}
{\mcitedefaultendpunct}{\mcitedefaultseppunct}\relax
\EndOfBibitem
\bibitem[Lee and Head-Gordon(2018)Lee, and Head-Gordon]{Lee2018}
Lee,~J.; Head-Gordon,~M. {Regularized Orbital-Optimized Second-Order
  M{\o}ller-Plesset Perturbation Theory: A Reliable Fifth-Order-Scaling
  Electron Correlation Model with Orbital Energy Dependent Regularizers}.
  \emph{J. Chem. Theory Comput.} \textbf{2018}, \emph{14}, 5203--5219\relax
\mciteBstWouldAddEndPuncttrue
\mciteSetBstMidEndSepPunct{\mcitedefaultmidpunct}
{\mcitedefaultendpunct}{\mcitedefaultseppunct}\relax
\EndOfBibitem
\bibitem[Lee and Head-Gordon(2019)Lee, and Head-Gordon]{Lee2019a}
Lee,~J.; Head-Gordon,~M. Distinguishing artificial and essential symmetry
  breaking in a single determinant: Approach and application to the \ce{C60},
  \ce{C36}, and \ce{C20} fullerenes. \emph{Phys. Chem. Chem. Phys.}
  \textbf{2019}, \emph{21}, 4763--4778\relax
\mciteBstWouldAddEndPuncttrue
\mciteSetBstMidEndSepPunct{\mcitedefaultmidpunct}
{\mcitedefaultendpunct}{\mcitedefaultseppunct}\relax
\EndOfBibitem
\bibitem[Deible and Jordan(2016)Deible, and Jordan]{deible2016exploration}
Deible,~M.~J.; Jordan,~K.~D. Exploration of Brueckner orbital trial wave
  functions in diffusion Monte Carlo calculations. \emph{Chem. Phys. Lett.}
  \textbf{2016}, \emph{644}, 117--120\relax
\mciteBstWouldAddEndPuncttrue
\mciteSetBstMidEndSepPunct{\mcitedefaultmidpunct}
{\mcitedefaultendpunct}{\mcitedefaultseppunct}\relax
\EndOfBibitem
\bibitem[Jim{\'{e}}nez-Hoyos \latin{et~al.}(2014)Jim{\'{e}}nez-Hoyos,
  Rodr{\'{i}}guez-Guzm{\'{a}}n, and Scuseria]{Jimenez-Hoyos2014}
Jim{\'{e}}nez-Hoyos,~C.~A.; Rodr{\'{i}}guez-Guzm{\'{a}}n,~R.; Scuseria,~G.~E.
  {Polyradical character and spin frustration in fullerene molecules: an ab
  initio non-collinear Hartree-Fock study.} \emph{J. Phys. Chem. A}
  \textbf{2014}, \emph{118}, 9925--40\relax
\mciteBstWouldAddEndPuncttrue
\mciteSetBstMidEndSepPunct{\mcitedefaultmidpunct}
{\mcitedefaultendpunct}{\mcitedefaultseppunct}\relax
\EndOfBibitem
\bibitem[Fowler \latin{et~al.}(1999)Fowler, Heine, Rogers, Sandall, Seifert,
  and Zerbetto]{Fowler1999}
Fowler,~P.; Heine,~T.; Rogers,~K.; Sandall,~J.; Seifert,~G.; Zerbetto,~F.
  {\ce{C36}, a hexavalent building block for fullerene compounds and solids}.
  \emph{Chem. Phys. Lett.} \textbf{1999}, \emph{300}, 369--378\relax
\mciteBstWouldAddEndPuncttrue
\mciteSetBstMidEndSepPunct{\mcitedefaultmidpunct}
{\mcitedefaultendpunct}{\mcitedefaultseppunct}\relax
\EndOfBibitem
\bibitem[Fowler \latin{et~al.}(1999)Fowler, Mitchell, and
  Zerbetto]{Fowler1999a}
Fowler,~P.~W.; Mitchell,~D.; Zerbetto,~F. \ce{C36}:~ The Best Fullerene for
  Covalent Bonding. \emph{J. Am. Chem. Soc.} \textbf{1999}, \emph{121},
  3218--3219\relax
\mciteBstWouldAddEndPuncttrue
\mciteSetBstMidEndSepPunct{\mcitedefaultmidpunct}
{\mcitedefaultendpunct}{\mcitedefaultseppunct}\relax
\EndOfBibitem
\bibitem[Jagadeesh and Chandrasekhar(1999)Jagadeesh, and
  Chandrasekhar]{Jagadeesh1999}
Jagadeesh,~M.~N.; Chandrasekhar,~J. {Computational studies on \ce{C36} and its
  dimer}. \emph{Chem. Phys. Lett.} \textbf{1999}, \emph{305}, 298--302\relax
\mciteBstWouldAddEndPuncttrue
\mciteSetBstMidEndSepPunct{\mcitedefaultmidpunct}
{\mcitedefaultendpunct}{\mcitedefaultseppunct}\relax
\EndOfBibitem
\bibitem[Aihara(1999)]{Aihara1999}
Aihara,~J.-i. {Weighted HOMO-LUMO energy separation as an index of kinetic
  stability for fullerenes}. \emph{Theor. Chem. Acc.} \textbf{1999},
  \emph{102}, 134--138\relax
\mciteBstWouldAddEndPuncttrue
\mciteSetBstMidEndSepPunct{\mcitedefaultmidpunct}
{\mcitedefaultendpunct}{\mcitedefaultseppunct}\relax
\EndOfBibitem
\bibitem[Yuan \latin{et~al.}(2000)Yuan, Yang, Deng, and Zhu]{Lan-FengYuan2000}
Yuan,~L.-F.; Yang,~J.; Deng,~K.; Zhu,~Q.-S. A First-Principles Study on the
  Structural and Electronic Properties of \ce{C36} Molecules. \emph{J. Phys.
  Chem. A} \textbf{2000}, \emph{104}, 6666--6671\relax
\mciteBstWouldAddEndPuncttrue
\mciteSetBstMidEndSepPunct{\mcitedefaultmidpunct}
{\mcitedefaultendpunct}{\mcitedefaultseppunct}\relax
\EndOfBibitem
\bibitem[Slanina \latin{et~al.}(2000)Slanina, Uhl\'{i}k, Zhao, and
  \={O}sawa]{Slanina2000}
Slanina,~Z.; Uhl\'{i}k,~F.; Zhao,~X.; \={O}sawa,~E. {Enthalpy-entropy interplay
  for \ce{C36} cages: B3LYP/6-31G* calculations}. \emph{J. Chem. Phys.}
  \textbf{2000}, \emph{113}, 4933\relax
\mciteBstWouldAddEndPuncttrue
\mciteSetBstMidEndSepPunct{\mcitedefaultmidpunct}
{\mcitedefaultendpunct}{\mcitedefaultseppunct}\relax
\EndOfBibitem
\bibitem[Ito \latin{et~al.}(2000)Ito, Monobe, Yoshii, and Tanaka]{Ito2000}
Ito,~A.; Monobe,~T.; Yoshii,~T.; Tanaka,~K. {Do \ce{C36} and \ce{C36H6}
  molecules have [36-D${}_\text{6h}$]fullerene structure?} \emph{Chem. Phys.
  Lett.} \textbf{2000}, \emph{328}, 32--38\relax
\mciteBstWouldAddEndPuncttrue
\mciteSetBstMidEndSepPunct{\mcitedefaultmidpunct}
{\mcitedefaultendpunct}{\mcitedefaultseppunct}\relax
\EndOfBibitem
\bibitem[Varganov \latin{et~al.}(2002)Varganov, Avramov, Ovchinnikov, and
  Gordon]{Varganov2002}
Varganov,~S.~A.; Avramov,~P.~V.; Ovchinnikov,~S.~G.; Gordon,~M.~S. {A study of
  the isomers of \ce{C36} fullerene using single and multireference MP2
  perturbation theory}. \emph{Chem. Phys. Lett.} \textbf{2002}, \emph{362},
  380--386\relax
\mciteBstWouldAddEndPuncttrue
\mciteSetBstMidEndSepPunct{\mcitedefaultmidpunct}
{\mcitedefaultendpunct}{\mcitedefaultseppunct}\relax
\EndOfBibitem
\bibitem[St{\"{u}}ck \latin{et~al.}(2011)St{\"{u}}ck, Baker, Zimmerman,
  Kurlancheek, and Head-Gordon]{Stuck2011}
St{\"{u}}ck,~D.; Baker,~T.~A.; Zimmerman,~P.; Kurlancheek,~W.; Head-Gordon,~M.
  {On the nature of electron correlation in \ce{C60}.} \emph{J. Chem. Phys.}
  \textbf{2011}, \emph{135}, 194306\relax
\mciteBstWouldAddEndPuncttrue
\mciteSetBstMidEndSepPunct{\mcitedefaultmidpunct}
{\mcitedefaultendpunct}{\mcitedefaultseppunct}\relax
\EndOfBibitem
\bibitem[Smith \latin{et~al.}(2017)Smith, Mussard, Holmes, and
  Sharma]{smith2017cheap}
Smith,~J.~E.; Mussard,~B.; Holmes,~A.~A.; Sharma,~S. Cheap and near exact
  CASSCF with large active spaces. \emph{J. Chem. Theory Comput.}
  \textbf{2017}, \emph{13}, 5468--5478\relax
\mciteBstWouldAddEndPuncttrue
\mciteSetBstMidEndSepPunct{\mcitedefaultmidpunct}
{\mcitedefaultendpunct}{\mcitedefaultseppunct}\relax
\EndOfBibitem
\bibitem[Levine \latin{et~al.}()Levine, Hait, Tubman, Lehtola, Whaley, and
  Head-Gordon]{levine2019casscf}
Levine,~D.~S.; Hait,~D.; Tubman,~N.~M.; Lehtola,~S.; Whaley,~K.~B.;
  Head-Gordon,~M. CASSCF with Extremely Large Active Spaces using the Adaptive
  Sampling Configuration Interaction Method. arXiv:1912.08379\relax
\mciteBstWouldAddEndPuncttrue
\mciteSetBstMidEndSepPunct{\mcitedefaultmidpunct}
{\mcitedefaultendpunct}{\mcitedefaultseppunct}\relax
\EndOfBibitem
\bibitem[Hohenstein \latin{et~al.}(2012)Hohenstein, Parrish, and
  Mart{\'{i}}nez]{Hohenstein2012}
Hohenstein,~E.~G.; Parrish,~R.~M.; Mart{\'{i}}nez,~T.~J. {Tensor
  hypercontraction density fitting. I. Quartic scaling second- and third-order
  M{\o}ller-Plesset perturbation theory}. \emph{J. Chem. Phys.} \textbf{2012},
  \emph{137}, 1085\relax
\mciteBstWouldAddEndPuncttrue
\mciteSetBstMidEndSepPunct{\mcitedefaultmidpunct}
{\mcitedefaultendpunct}{\mcitedefaultseppunct}\relax
\EndOfBibitem
\bibitem[Lee \latin{et~al.}(2019)Lee, Lin, and Head-Gordon]{lee_thc}
Lee,~J.; Lin,~L.; Head-Gordon,~M. Systematically Improvable Tensor
  Hypercontraction: Interpolative Separable Density-Fitting for Molecules
  Applied to Exact Exchange, Second- and Third-Order
  M{\o}ller{\textendash}Plesset Perturbation Theory. \emph{J. Chem. Theory
  Comput.} \textbf{2019}, \emph{16}, 243--263\relax
\mciteBstWouldAddEndPuncttrue
\mciteSetBstMidEndSepPunct{\mcitedefaultmidpunct}
{\mcitedefaultendpunct}{\mcitedefaultseppunct}\relax
\EndOfBibitem
\bibitem[Lee \latin{et~al.}(2019)Lee, Bertels, Small, and
  Head-Gordon]{lee2019kohn}
Lee,~J.; Bertels,~L.~W.; Small,~D.~W.; Head-Gordon,~M. Kohn-sham density
  functional theory with complex, spin-restricted orbitals: Accessing a new
  class of densities without the symmetry dilemma. \emph{Phys. Rev. Lett.}
  \textbf{2019}, \emph{123}, 113001\relax
\mciteBstWouldAddEndPuncttrue
\mciteSetBstMidEndSepPunct{\mcitedefaultmidpunct}
{\mcitedefaultendpunct}{\mcitedefaultseppunct}\relax
\EndOfBibitem
\bibitem[Small \latin{et~al.}(2015)Small, Sundstrom, and
  Head-Gordon]{Small2015a}
Small,~D.~W.; Sundstrom,~E.~J.; Head-Gordon,~M. {Restricted Hartree Fock using
  complex-valued orbitals: a long-known but neglected tool in electronic
  structure theory.} \emph{J. Chem. Phys.} \textbf{2015}, \emph{142},
  024104\relax
\mciteBstWouldAddEndPuncttrue
\mciteSetBstMidEndSepPunct{\mcitedefaultmidpunct}
{\mcitedefaultendpunct}{\mcitedefaultseppunct}\relax
\EndOfBibitem
\bibitem[Dunning(1989)]{Dunning1989}
Dunning,~T.~H. {Gaussian basis sets for use in correlated molecular
  calculations. I. The atoms boron through neon and hydrogen}. \emph{J. Chem.
  Phys.} \textbf{1989}, \emph{90}, 1007--1023\relax
\mciteBstWouldAddEndPuncttrue
\mciteSetBstMidEndSepPunct{\mcitedefaultmidpunct}
{\mcitedefaultendpunct}{\mcitedefaultseppunct}\relax
\EndOfBibitem
\bibitem[Helgaker \latin{et~al.}(1997)Helgaker, Klopper, Koch, and
  Noga]{helgaker1997basis}
Helgaker,~T.; Klopper,~W.; Koch,~H.; Noga,~J. Basis-set convergence of
  correlated calculations on water. \emph{J. Chem. Phys.} \textbf{1997},
  \emph{106}, 9639--9646\relax
\mciteBstWouldAddEndPuncttrue
\mciteSetBstMidEndSepPunct{\mcitedefaultmidpunct}
{\mcitedefaultendpunct}{\mcitedefaultseppunct}\relax
\EndOfBibitem
\bibitem[Hehre \latin{et~al.}(1972)Hehre, Ditchfield, and Pople]{Hehre1972}
Hehre,~W.~J.; Ditchfield,~R.; Pople,~J.~A. Self{\textemdash}Consistent
  Molecular Orbital Methods. {XII}. Further Extensions of
  Gaussian{\textemdash}Type Basis Sets for Use in Molecular Orbital Studies of
  Organic Molecules. \emph{J. Chem. Phys.} \textbf{1972}, \emph{56},
  2257--2261\relax
\mciteBstWouldAddEndPuncttrue
\mciteSetBstMidEndSepPunct{\mcitedefaultmidpunct}
{\mcitedefaultendpunct}{\mcitedefaultseppunct}\relax
\EndOfBibitem
\bibitem[Yamaguchi \latin{et~al.}(1988)Yamaguchi, Jensen, Dorigo, and
  Houk]{Yamaguchi1988}
Yamaguchi,~K.; Jensen,~F.; Dorigo,~A.; Houk,~K.~N. {A spin correction procedure
  for unrestricted Hartree-Fock and M{\o}ller-Plesset wavefunctions for singlet
  diradicals and polyradicals}. \emph{Chem. Phys. Lett.} \textbf{1988},
  \emph{149}, 537--542\relax
\mciteBstWouldAddEndPuncttrue
\mciteSetBstMidEndSepPunct{\mcitedefaultmidpunct}
{\mcitedefaultendpunct}{\mcitedefaultseppunct}\relax
\EndOfBibitem
\bibitem[Groenhof \latin{et~al.}(2005)Groenhof, Swart, Ehlers, and
  Lammertsma]{Groenhof2005}
Groenhof,~A.~R.; Swart,~M.; Ehlers,~A.~W.; Lammertsma,~K. {Electronic ground
  states of iron porphyrin and of the first species in the catalytic reaction
  cycle of cytochrome P450s}. \emph{J. Phys. Chem. A} \textbf{2005},
  \emph{109}, 3411--3417\relax
\mciteBstWouldAddEndPuncttrue
\mciteSetBstMidEndSepPunct{\mcitedefaultmidpunct}
{\mcitedefaultendpunct}{\mcitedefaultseppunct}\relax
\EndOfBibitem
\bibitem[Hori and Kitagawa(1980)Hori, and Kitagawa]{hori1980iron}
Hori,~H.; Kitagawa,~T. Iron-ligand stretching band in the resonance Raman
  spectra of ferrous iron porphyrin derivatives. Importance as a probe band for
  quaternary structure of hemoglobin. \emph{J. Am. Chem. Soc.} \textbf{1980},
  \emph{102}, 3608--3613\relax
\mciteBstWouldAddEndPuncttrue
\mciteSetBstMidEndSepPunct{\mcitedefaultmidpunct}
{\mcitedefaultendpunct}{\mcitedefaultseppunct}\relax
\EndOfBibitem
\bibitem[Matsuzawa \latin{et~al.}(1995)Matsuzawa, Ata, and
  Dixon]{Matsuzawa1995}
Matsuzawa,~N.; Ata,~M.; Dixon,~D.~A. {Density functional theory prediction of
  the second-order hyperpolarizability of metalloporphines}. \emph{J. Phys.
  Chem.} \textbf{1995}, \emph{99}, 7698--7706\relax
\mciteBstWouldAddEndPuncttrue
\mciteSetBstMidEndSepPunct{\mcitedefaultmidpunct}
{\mcitedefaultendpunct}{\mcitedefaultseppunct}\relax
\EndOfBibitem
\bibitem[Rovira \latin{et~al.}(1997)Rovira, Kunc, Hutter, Ballone, and
  Parrinello]{Rovira1997}
Rovira,~C.; Kunc,~K.; Hutter,~J.; Ballone,~P.; Parrinello,~M. {Equilibrium
  geometries and electronic structure of iron-porphyrin complexes: A density
  functional study}. \emph{J. Phys. Chem. A} \textbf{1997}, \emph{101},
  8914--8925\relax
\mciteBstWouldAddEndPuncttrue
\mciteSetBstMidEndSepPunct{\mcitedefaultmidpunct}
{\mcitedefaultendpunct}{\mcitedefaultseppunct}\relax
\EndOfBibitem
\bibitem[Radon{\'n} and Pierloot(2008)Radon{\'n}, and Pierloot]{Radon2008}
Radon{\'n},~M.; Pierloot,~K. {Binding of CO, NO, and O 2 to Heme by Density
  Functional and Multireference ab Initio Calculations}. \emph{J. Phys. Chem.
  A} \textbf{2008}, \emph{112}, 11824--11832\relax
\mciteBstWouldAddEndPuncttrue
\mciteSetBstMidEndSepPunct{\mcitedefaultmidpunct}
{\mcitedefaultendpunct}{\mcitedefaultseppunct}\relax
\EndOfBibitem
\bibitem[Rado{\'{n}}(2014)]{Radon2014}
Rado{\'{n}},~M. {Spin-State Energetics of Heme-Related Models from DFT and
  Coupled Cluster Calculations}. \emph{J. Chem. Theory Comput.} \textbf{2014},
  \emph{10}, 2306--2321\relax
\mciteBstWouldAddEndPuncttrue
\mciteSetBstMidEndSepPunct{\mcitedefaultmidpunct}
{\mcitedefaultendpunct}{\mcitedefaultseppunct}\relax
\EndOfBibitem
\bibitem[Choe \latin{et~al.}(1999)Choe, Nakajima, Hirao, and Lindh]{Choe1999}
Choe,~Y.~K.; Nakajima,~T.; Hirao,~K.; Lindh,~R. {Theoretical study of the
  electronic ground state of iron(II) porphine. II}. \emph{J. Chem. Phys.}
  \textbf{1999}, \emph{111}, 3837--3845\relax
\mciteBstWouldAddEndPuncttrue
\mciteSetBstMidEndSepPunct{\mcitedefaultmidpunct}
{\mcitedefaultendpunct}{\mcitedefaultseppunct}\relax
\EndOfBibitem
\bibitem[Pierloot(2003)]{Pierloot2003}
Pierloot,~K. {The CASPT2 method in inorganic electronic spectroscopy: From
  ionic transition metal to covalent actinide complexes}. \emph{Mol. Phys.}
  \textbf{2003}, \emph{101}, 2083--2094\relax
\mciteBstWouldAddEndPuncttrue
\mciteSetBstMidEndSepPunct{\mcitedefaultmidpunct}
{\mcitedefaultendpunct}{\mcitedefaultseppunct}\relax
\EndOfBibitem
\bibitem[Phung \latin{et~al.}(2018)Phung, Feldt, Harvey, and
  Pierloot]{Phung2018}
Phung,~Q.~M.; Feldt,~M.; Harvey,~J.~N.; Pierloot,~K. {Toward Highly Accurate
  Spin State Energetics in First-Row Transition Metal Complexes: A Combined
  CASPT2/CC Approach}. \emph{J. Chem. Theory Comput.} \textbf{2018}, \emph{14},
  2446--2455\relax
\mciteBstWouldAddEndPuncttrue
\mciteSetBstMidEndSepPunct{\mcitedefaultmidpunct}
{\mcitedefaultendpunct}{\mcitedefaultseppunct}\relax
\EndOfBibitem
\bibitem[Vancoillie \latin{et~al.}(2011)Vancoillie, Zhao, Tran, Hendrickx, and
  Pierloot]{Vancoillie2011}
Vancoillie,~S.; Zhao,~H.; Tran,~V.~T.; Hendrickx,~M.~F.; Pierloot,~K.
  {Multiconfigurational second-order perturbation theory restricted active
  space (RASPT2) studies on mononuclear first-row transition-metal systems}.
  \emph{J. Chem. Theory Comput.} \textbf{2011}, \emph{7}, 3961--3977\relax
\mciteBstWouldAddEndPuncttrue
\mciteSetBstMidEndSepPunct{\mcitedefaultmidpunct}
{\mcitedefaultendpunct}{\mcitedefaultseppunct}\relax
\EndOfBibitem
\bibitem[Olivares-Amaya \latin{et~al.}(2015)Olivares-Amaya, Hu, Nakatani,
  Sharma, Yang, and Chan]{Olivares-Amaya2015}
Olivares-Amaya,~R.; Hu,~W.; Nakatani,~N.; Sharma,~S.; Yang,~J.; Chan,~G. K.~L.
  {The ab-initio density matrix renormalization group in practice}. \emph{J.
  Chem. Phys.} \textbf{2015}, \emph{142}\relax
\mciteBstWouldAddEndPuncttrue
\mciteSetBstMidEndSepPunct{\mcitedefaultmidpunct}
{\mcitedefaultendpunct}{\mcitedefaultseppunct}\relax
\EndOfBibitem
\bibitem[Phung \latin{et~al.}(2016)Phung, Wouters, and Pierloot]{Phung2016}
Phung,~Q.~M.; Wouters,~S.; Pierloot,~K. {Cumulant Approximated Second-Order
  Perturbation Theory Based on the Density Matrix Renormalization Group for
  Transition Metal Complexes: A Benchmark Study}. \emph{J. Chem. Theory
  Comput.} \textbf{2016}, \emph{12}, 4352--4361\relax
\mciteBstWouldAddEndPuncttrue
\mciteSetBstMidEndSepPunct{\mcitedefaultmidpunct}
{\mcitedefaultendpunct}{\mcitedefaultseppunct}\relax
\EndOfBibitem
\bibitem[{Li Manni} and Alavi(2018){Li Manni}, and Alavi]{LiManni2018}
{Li Manni},~G.; Alavi,~A. {Understanding the Mechanism Stabilizing Intermediate
  Spin States in Fe(II)-Porphyrin}. \emph{J. Phys. Chem. A} \textbf{2018},
  \emph{122}, 4935--4947\relax
\mciteBstWouldAddEndPuncttrue
\mciteSetBstMidEndSepPunct{\mcitedefaultmidpunct}
{\mcitedefaultendpunct}{\mcitedefaultseppunct}\relax
\EndOfBibitem
\bibitem[{Li Manni} \latin{et~al.}(2019){Li Manni}, Kats, Tew, and
  Alavi]{LiManni2019}
{Li Manni},~G.; Kats,~D.; Tew,~D.~P.; Alavi,~A. {Role of Valence and Semicore
  Electron Correlation on Spin Gaps in Fe(II)-Porphyrins}. \emph{J. Chem.
  Theory Comput.} \textbf{2019}, \emph{15}, 1492--1497\relax
\mciteBstWouldAddEndPuncttrue
\mciteSetBstMidEndSepPunct{\mcitedefaultmidpunct}
{\mcitedefaultendpunct}{\mcitedefaultseppunct}\relax
\EndOfBibitem
\bibitem[Zhou \latin{et~al.}(2019)Zhou, Gagliardi, and Truhlar]{Zhou2019}
Zhou,~C.; Gagliardi,~L.; Truhlar,~D.~G. {Multiconfiguration Pair-Density
  Functional Theory for Iron Porphyrin with CAS, RAS, and DMRG Active Spaces}.
  \emph{J. Phys. Chem. A} \textbf{2019}, \emph{123}, 3389--3394\relax
\mciteBstWouldAddEndPuncttrue
\mciteSetBstMidEndSepPunct{\mcitedefaultmidpunct}
{\mcitedefaultendpunct}{\mcitedefaultseppunct}\relax
\EndOfBibitem
\bibitem[Collman \latin{et~al.}(1975)Collman, Hoard, Kim, Lang, and
  Reed]{collman1975synthesis}
Collman,~J.~P.; Hoard,~J.; Kim,~N.; Lang,~G.; Reed,~C.~A. Synthesis,
  stereochemistry, and structure-related properties of. alpha.,. beta.,.
  gamma.,. delta.-tetraphenylporphinatoiron (II). \emph{J. Am. Chem. Soc.}
  \textbf{1975}, \emph{97}, 2676--2681\relax
\mciteBstWouldAddEndPuncttrue
\mciteSetBstMidEndSepPunct{\mcitedefaultmidpunct}
{\mcitedefaultendpunct}{\mcitedefaultseppunct}\relax
\EndOfBibitem
\bibitem[Dolphin \latin{et~al.}(1976)Dolphin, Sams, Tsin, and
  Wong]{dolphin1976synthesis}
Dolphin,~D.; Sams,~J.~R.; Tsin,~T.~B.; Wong,~K.~L. Synthesis and Moessbauer
  spectra of octaethylporphyrin ferrous complexes. \emph{J. Am. Chem. Soc.}
  \textbf{1976}, \emph{98}, 6970--6975\relax
\mciteBstWouldAddEndPuncttrue
\mciteSetBstMidEndSepPunct{\mcitedefaultmidpunct}
{\mcitedefaultendpunct}{\mcitedefaultseppunct}\relax
\EndOfBibitem
\bibitem[Goff \latin{et~al.}(1977)Goff, La~Mar, and Reed]{goff1977nuclear}
Goff,~H.; La~Mar,~G.~N.; Reed,~C.~A. Nuclear magnetic resonance investigation
  of magnetic and electronic properties of" intermediate spin" ferrous
  porphyrin complexes. \emph{J. Am. Chem. Soc.} \textbf{1977}, \emph{99},
  3641--3646\relax
\mciteBstWouldAddEndPuncttrue
\mciteSetBstMidEndSepPunct{\mcitedefaultmidpunct}
{\mcitedefaultendpunct}{\mcitedefaultseppunct}\relax
\EndOfBibitem
\bibitem[Lang \latin{et~al.}(1978)Lang, Spartalian, Reed, and
  Collman]{lang1978mossbauer}
Lang,~G.; Spartalian,~K.; Reed,~C.~A.; Collman,~J.~P. M{\"o}ssbauer effect
  study of the magnetic properties of S= 1 ferrous tetraphenylporphyrin.
  \emph{J. Chem. Phys.} \textbf{1978}, \emph{69}, 5424--5427\relax
\mciteBstWouldAddEndPuncttrue
\mciteSetBstMidEndSepPunct{\mcitedefaultmidpunct}
{\mcitedefaultendpunct}{\mcitedefaultseppunct}\relax
\EndOfBibitem
\bibitem[Kitagawa and Teraoka(1979)Kitagawa, and
  Teraoka]{kitagawa1979resonance}
Kitagawa,~T.; Teraoka,~J. The resonance Raman spectra of intermediate-spin
  ferrous porphyrin. \emph{Chem. Phys. Lett.} \textbf{1979}, \emph{63},
  443--446\relax
\mciteBstWouldAddEndPuncttrue
\mciteSetBstMidEndSepPunct{\mcitedefaultmidpunct}
{\mcitedefaultendpunct}{\mcitedefaultseppunct}\relax
\EndOfBibitem
\bibitem[Boyd \latin{et~al.}(1979)Boyd, Buckingham, McMeeking, and
  Mitra]{boyd1979paramagnetic}
Boyd,~P.~D.; Buckingham,~D.~A.; McMeeking,~R.~F.; Mitra,~S. Paramagnetic
  anisotropy, average magnetic susceptibility, and electronic structure of
  intermediate-spin S= 1 (5, 10, 15, 20-tetraphenylporphyrin) iron (II).
  \emph{Inorg. Chem.} \textbf{1979}, \emph{18}, 3585--3591\relax
\mciteBstWouldAddEndPuncttrue
\mciteSetBstMidEndSepPunct{\mcitedefaultmidpunct}
{\mcitedefaultendpunct}{\mcitedefaultseppunct}\relax
\EndOfBibitem
\bibitem[Mispelter \latin{et~al.}(1980)Mispelter, Momenteau, and
  Lhoste]{mispelter1980proton}
Mispelter,~J.; Momenteau,~M.; Lhoste,~J. Proton magnetic resonance
  characterization of the intermediate (S= 1) spin state of ferrous porphyrins.
  \emph{J. Chem. Phys.} \textbf{1980}, \emph{72}, 1003--1012\relax
\mciteBstWouldAddEndPuncttrue
\mciteSetBstMidEndSepPunct{\mcitedefaultmidpunct}
{\mcitedefaultendpunct}{\mcitedefaultseppunct}\relax
\EndOfBibitem
\bibitem[Strauss \latin{et~al.}(1985)Strauss, Silver, Long, Thompson, Hudgens,
  Spartalian, and Ibers]{strauss1985comparison}
Strauss,~S.~H.; Silver,~M.~E.; Long,~K.~M.; Thompson,~R.~G.; Hudgens,~R.~A.;
  Spartalian,~K.; Ibers,~J.~A. Comparison of the molecular and electronic
  structures of (2, 3, 7, 8, 12, 13, 17, 18-octaethylporphyrinato) iron (II)
  and (trans-7, 8-dihydro-2, 3, 7, 8, 12, 13, 17, 18-octaethylporphyrinato)
  iron (II). \emph{J. Am. Chem. Soc.} \textbf{1985}, \emph{107},
  4207--4215\relax
\mciteBstWouldAddEndPuncttrue
\mciteSetBstMidEndSepPunct{\mcitedefaultmidpunct}
{\mcitedefaultendpunct}{\mcitedefaultseppunct}\relax
\EndOfBibitem
\bibitem[Pierloot \latin{et~al.}(2017)Pierloot, Phung, and
  Domingo]{Pierloot2017}
Pierloot,~K.; Phung,~Q.~M.; Domingo,~A. {Spin State Energetics in First-Row
  Transition Metal Complexes: Contribution of (3s3p) Correlation and Its
  Description by Second-Order Perturbation Theory}. \emph{J. Chem. Theory
  Comput.} \textbf{2017}, \emph{13}, 537--553\relax
\mciteBstWouldAddEndPuncttrue
\mciteSetBstMidEndSepPunct{\mcitedefaultmidpunct}
{\mcitedefaultendpunct}{\mcitedefaultseppunct}\relax
\EndOfBibitem
\bibitem[Douglas and Kroll(1974)Douglas, and Kroll]{douglas1974quantum}
Douglas,~M.; Kroll,~N.~M. Quantum electrodynamical corrections to the fine
  structure of helium. \emph{Ann. Phys.} \textbf{1974}, \emph{82},
  89--155\relax
\mciteBstWouldAddEndPuncttrue
\mciteSetBstMidEndSepPunct{\mcitedefaultmidpunct}
{\mcitedefaultendpunct}{\mcitedefaultseppunct}\relax
\EndOfBibitem
\bibitem[Widmark \latin{et~al.}(1990)Widmark, Malmqvist, and
  Roos]{widmark1990density}
Widmark,~P.-O.; Malmqvist,~P.-{\AA}.; Roos,~B.~O. Density matrix averaged
  atomic natural orbital (ANO) basis sets for correlated molecular wave
  functions. \emph{Theor. Chim. Acta.} \textbf{1990}, \emph{77}, 291--306\relax
\mciteBstWouldAddEndPuncttrue
\mciteSetBstMidEndSepPunct{\mcitedefaultmidpunct}
{\mcitedefaultendpunct}{\mcitedefaultseppunct}\relax
\EndOfBibitem
\bibitem[Roos \latin{et~al.}(2004)Roos, Lindh, Malmqvist, Veryazov, and
  Widmark]{roos2004main}
Roos,~B.~O.; Lindh,~R.; Malmqvist,~P.-{\AA}.; Veryazov,~V.; Widmark,~P.-O. Main
  group atoms and dimers studied with a new relativistic ANO basis set.
  \emph{J. Phys. Chem. A} \textbf{2004}, \emph{108}, 2851--2858\relax
\mciteBstWouldAddEndPuncttrue
\mciteSetBstMidEndSepPunct{\mcitedefaultmidpunct}
{\mcitedefaultendpunct}{\mcitedefaultseppunct}\relax
\EndOfBibitem
\bibitem[Roos \latin{et~al.}(2005)Roos, Lindh, Malmqvist, Veryazov, and
  Widmark]{roos2005new}
Roos,~B.~O.; Lindh,~R.; Malmqvist,~P.-{\AA}.; Veryazov,~V.; Widmark,~P.-O. New
  relativistic ANO basis sets for transition metal atoms. \emph{J. Phys. Chem.
  A} \textbf{2005}, \emph{109}, 6575--6579\relax
\mciteBstWouldAddEndPuncttrue
\mciteSetBstMidEndSepPunct{\mcitedefaultmidpunct}
{\mcitedefaultendpunct}{\mcitedefaultseppunct}\relax
\EndOfBibitem
\bibitem[Bertels \latin{et~al.}(2019)Bertels, Lee, and
  Head-Gordon]{bertels2019third}
Bertels,~L.~W.; Lee,~J.; Head-Gordon,~M. Third-Order M{\o}ller-Plesset
  Perturbation Theory Made Useful? Choice of Orbitals and Scaling Greatly
  Improves Accuracy for Thermochemistry, Kinetics, and Intermolecular
  Interactions. \emph{J. Phys. Chem. Lett.} \textbf{2019}, \emph{10},
  4170--4176\relax
\mciteBstWouldAddEndPuncttrue
\mciteSetBstMidEndSepPunct{\mcitedefaultmidpunct}
{\mcitedefaultendpunct}{\mcitedefaultseppunct}\relax
\EndOfBibitem
\bibitem[Shao \latin{et~al.}(2015)Shao, Gan, Epifanovsky, Gilbert, Wormit,
  Kussmann, Lange, Behn, Deng, Feng, Ghosh, Goldey, Horn, Jacobson, Kaliman,
  Khaliullin, Ku{\' s}, Landau, Liu, Proynov, Rhee, Richard, Rohrdanz, Steele,
  Sundstrom, Woodcock, Zimmerman, Zuev, Albrecht, Alguire, Austin, Beran,
  Bernard, Berquist, Brandhorst, Bravaya, Brown, Casanova, Chang, Chen, Chien,
  Closser, Crittenden, Diedenhofen, Distasio, Do, Dutoi, Edgar, Fatehi,
  Fusti-Molnar, Ghysels, Golubeva-Zadorozhnaya, Gomes, Hanson-Heine, Harbach,
  Hauser, Hohenstein, Holden, Jagau, Ji, Kaduk, Khistyaev, Kim, Kim, King,
  Klunzinger, Kosenkov, Kowalczyk, Krauter, Lao, Laurent, Lawler, Levchenko,
  Lin, Liu, Livshits, Lochan, Luenser, Manohar, Manzer, Mao, Mardirossian,
  Marenich, Maurer, Mayhall, Neuscamman, Oana, Olivares-Amaya, Oneill,
  Parkhill, Perrine, Peverati, Prociuk, Rehn, Rosta, Russ, Sharada, Sharma,
  Small, Sodt, Stein, St{\"{u}}ck, Su, Thom, Tsuchimochi, Vanovschi, Vogt,
  Vydrov, Wang, Watson, Wenzel, White, Williams, Yang, Yeganeh, Yost, You,
  Zhang, Zhang, Zhao, Brooks, Chan, Chipman, Cramer, Goddard, Gordon, Hehre,
  Klamt, Schaefer, Schmidt, Sherrill, Truhlar, Warshel, Xu, Aspuru-Guzik, Baer,
  Bell, Besley, Chai, Dreuw, Dunietz, Furlani, Gwaltney, Hsu, Jung, Kong,
  Lambrecht, Liang, Ochsenfeld, Rassolov, Slipchenko, Subotnik, {Van Voorhis},
  Herbert, Krylov, Gill, and Head-Gordon]{Shao2015}
Shao,~Y.; Gan,~Z.; Epifanovsky,~E.; Gilbert,~A.~T.; Wormit,~M.; Kussmann,~J.;
  Lange,~A.~W.; Behn,~A.; Deng,~J.; Feng,~X.; Ghosh,~D.; Goldey,~M.;
  Horn,~P.~R.; Jacobson,~L.~D.; Kaliman,~I.; Khaliullin,~R.~Z.; Ku{\' s},~T.;
  Landau,~A.; Liu,~J.; Proynov,~E.~I.; Rhee,~Y.~M.; Richard,~R.~M.;
  Rohrdanz,~M.~A.; Steele,~R.~P.; Sundstrom,~E.~J.; Woodcock,~H.~L.;
  Zimmerman,~P.~M.; Zuev,~D.; Albrecht,~B.; Alguire,~E.; Austin,~B.;
  Beran,~G.~J.; Bernard,~Y.~A.; Berquist,~E.; Brandhorst,~K.; Bravaya,~K.~B.;
  Brown,~S.~T.; Casanova,~D.; Chang,~C.~M.; Chen,~Y.; Chien,~S.~H.;
  Closser,~K.~D.; Crittenden,~D.~L.; Diedenhofen,~M.; Distasio,~R.~A.; Do,~H.;
  Dutoi,~A.~D.; Edgar,~R.~G.; Fatehi,~S.; Fusti-Molnar,~L.; Ghysels,~A.;
  Golubeva-Zadorozhnaya,~A.; Gomes,~J.; Hanson-Heine,~M.~W.; Harbach,~P.~H.;
  Hauser,~A.~W.; Hohenstein,~E.~G.; Holden,~Z.~C.; Jagau,~T.~C.; Ji,~H.;
  Kaduk,~B.; Khistyaev,~K.; Kim,~J.; Kim,~J.; King,~R.~A.; Klunzinger,~P.;
  Kosenkov,~D.; Kowalczyk,~T.; Krauter,~C.~M.; Lao,~K.~U.; Laurent,~A.~D.;
  Lawler,~K.~V.; Levchenko,~S.~V.; Lin,~C.~Y.; Liu,~F.; Livshits,~E.;
  Lochan,~R.~C.; Luenser,~A.; Manohar,~P.; Manzer,~S.~F.; Mao,~S.~P.;
  Mardirossian,~N.; Marenich,~A.~V.; Maurer,~S.~A.; Mayhall,~N.~J.;
  Neuscamman,~E.; Oana,~C.~M.; Olivares-Amaya,~R.; Oneill,~D.~P.;
  Parkhill,~J.~A.; Perrine,~T.~M.; Peverati,~R.; Prociuk,~A.; Rehn,~D.~R.;
  Rosta,~E.; Russ,~N.~J.; Sharada,~S.~M.; Sharma,~S.; Small,~D.~W.; Sodt,~A.;
  Stein,~T.; St{\"{u}}ck,~D.; Su,~Y.~C.; Thom,~A.~J.; Tsuchimochi,~T.;
  Vanovschi,~V.; Vogt,~L.; Vydrov,~O.; Wang,~T.; Watson,~M.~A.; Wenzel,~J.;
  White,~A.; Williams,~C.~F.; Yang,~J.; Yeganeh,~S.; Yost,~S.~R.; You,~Z.~Q.;
  Zhang,~I.~Y.; Zhang,~X.; Zhao,~Y.; Brooks,~B.~R.; Chan,~G.~K.;
  Chipman,~D.~M.; Cramer,~C.~J.; Goddard,~W.~A.; Gordon,~M.~S.; Hehre,~W.~J.;
  Klamt,~A.; Schaefer,~H.~F.; Schmidt,~M.~W.; Sherrill,~C.~D.; Truhlar,~D.~G.;
  Warshel,~A.; Xu,~X.; Aspuru-Guzik,~A.; Baer,~R.; Bell,~A.~T.; Besley,~N.~A.;
  Chai,~J.~D.; Dreuw,~A.; Dunietz,~B.~D.; Furlani,~T.~R.; Gwaltney,~S.~R.;
  Hsu,~C.~P.; Jung,~Y.; Kong,~J.; Lambrecht,~D.~S.; Liang,~W.; Ochsenfeld,~C.;
  Rassolov,~V.~A.; Slipchenko,~L.~V.; Subotnik,~J.~E.; {Van Voorhis},~T.;
  Herbert,~J.~M.; Krylov,~A.~I.; Gill,~P.~M.; Head-Gordon,~M. {Advances in
  molecular quantum chemistry contained in the Q-Chem 4 program package}.
  \emph{Mol. Phys.} \textbf{2015}, \emph{113}, 184--215\relax
\mciteBstWouldAddEndPuncttrue
\mciteSetBstMidEndSepPunct{\mcitedefaultmidpunct}
{\mcitedefaultendpunct}{\mcitedefaultseppunct}\relax
\EndOfBibitem
\bibitem[Kim \latin{et~al.}(2018)Kim, Baczewski, Beaudet, Benali, Bennett,
  Berrill, Blunt, Borda, Casula, Ceperley, Chiesa, Clark, III, Delaney, Dewing,
  Esler, Hao, Heinonen, Kent, Krogel, Kyl{\"a}np{\"a}{\"a}, Li, Lopez, Luo,
  Malone, Martin, Mathuriya, McMinis, Melton, Mitas, Morales, Neuscamman,
  Parker, Flores, Romero, Rubenstein, Shea, Shin, Shulenburger, Tillack,
  Townsend, Tubman, Goetz, Vincent, Yang, Yang, Zhang, and Zhao]{qmcpack}
Kim,~J.; Baczewski,~A.~T.; Beaudet,~T.~D.; Benali,~A.; Bennett,~M.~C.;
  Berrill,~M.~A.; Blunt,~N.~S.; Borda,~E. J.~L.; Casula,~M.; Ceperley,~D.~M.;
  Chiesa,~S.; Clark,~B.~K.; III,~R. C.~C.; Delaney,~K.~T.; Dewing,~M.;
  Esler,~K.~P.; Hao,~H.; Heinonen,~O.; Kent,~P. R.~C.; Krogel,~J.~T.;
  Kyl{\"a}np{\"a}{\"a},~I.; Li,~Y.~W.; Lopez,~M.~G.; Luo,~Y.; Malone,~F.~D.;
  Martin,~R.~M.; Mathuriya,~A.; McMinis,~J.; Melton,~C.~A.; Mitas,~L.;
  Morales,~M.~A.; Neuscamman,~E.; Parker,~W.~D.; Flores,~S. D.~P.;
  Romero,~N.~A.; Rubenstein,~B.~M.; Shea,~J. A.~R.; Shin,~H.; Shulenburger,~L.;
  Tillack,~A.~F.; Townsend,~J.~P.; Tubman,~N.~M.; Goetz,~B. V.~D.;
  Vincent,~J.~E.; Yang,~D.~C.; Yang,~Y.; Zhang,~S.; Zhao,~L. QMCPACK : an open
  source ab initio quantum Monte Carlo package for the electronic structure of
  atoms, molecules and solids. \emph{J. Phys.: Cond. Mat.} \textbf{2018},
  \emph{30}, 195901\relax
\mciteBstWouldAddEndPuncttrue
\mciteSetBstMidEndSepPunct{\mcitedefaultmidpunct}
{\mcitedefaultendpunct}{\mcitedefaultseppunct}\relax
\EndOfBibitem
\bibitem[Kent \latin{et~al.}(2020)Kent, Annaberdiyev, Benali, Bennett, Borda,
  Doak, Jordan, Krogel, Kylanpaa, Lee, \latin{et~al.} others]{kent2020qmcpack}
Kent,~P.; Annaberdiyev,~A.; Benali,~A.; Bennett,~M.~C.; Borda,~E. J.~L.;
  Doak,~P.; Jordan,~K.~D.; Krogel,~J.~T.; Kylanpaa,~I.; Lee,~J., \latin{et~al.}
   QMCPACK: Advances in the development, efficiency, and application of
  auxiliary field and real-space variational and diffusion Quantum Monte Carlo.
  \emph{arXiv:2003.01831} \textbf{2020}, \relax
\mciteBstWouldAddEndPunctfalse
\mciteSetBstMidEndSepPunct{\mcitedefaultmidpunct}
{}{\mcitedefaultseppunct}\relax
\EndOfBibitem
\bibitem[Sun \latin{et~al.}(2017)Sun, Berkelbach, Blunt, Booth, Guo, Li, Liu,
  McClain, Sayfutyarova, Sharma, Wouters, and Chan]{PYSCF}
Sun,~Q.; Berkelbach,~T.~C.; Blunt,~N.~S.; Booth,~G.~H.; Guo,~S.; Li,~Z.;
  Liu,~J.; McClain,~J.~D.; Sayfutyarova,~E.~R.; Sharma,~S.; Wouters,~S.;
  Chan,~G. K.~L. PySCF: the Python-based simulations of chemistry framework.
  \emph{WIREs Comput. Mol. Sci.} \textbf{2017}, \emph{8}, e1340\relax
\mciteBstWouldAddEndPuncttrue
\mciteSetBstMidEndSepPunct{\mcitedefaultmidpunct}
{\mcitedefaultendpunct}{\mcitedefaultseppunct}\relax
\EndOfBibitem
\end{mcitethebibliography}
\bibliographystyle{achemso}
\end{document}